%Paper: 9201056
%From: ZWIEBACH%IASSNS.BITNET@pucc.princeton.edu
%Date: Mon, 27 Jan 92 21:30 EST

\input phyzzx

%%%%%%%%%%%%%%%%%%%%%%%%%%%%%%%%%%%%%%%%%%%%%%%%%%%%%%%%%%%%%%%%
% This will make your PHYZZX pagesize wider and longer
% It is OPTIONAL. It redefines the papers macro
%
%\catcode`\@=11 % This allows us to modify PLAIN macros.
%
%\def\papers{\papersize\headline=\paperheadline\footline=\paperfootline}
%
%\def\papersize{\hsize=40pc \vsize=53pc \hoffset=0pc \voffset=1pc
%   \advance\hoffset by\HOFFSET \advance\voffset by\VOFFSET
%   \pagebottomfiller=0pc
%   \skip\footins=\bigskipamount \normalspace }
%
%\catcode`\@=12 % at signs are no longer letters
%
\papers
%%%%%%%%%%%%%%%%%%%%%%%%%%%%%%%%%%%%%%%%%%%%%%%%%%%%%%%%%%%%%%%%
\overfullrule=0pt
\def\hat{\widehat}

\def\bar{\overline}
\def\IR{{\hbox{{\rm I}\kern-.2em\hbox{\rm R}}}}
\def\IB{{\hbox{{\rm I}\kern-.2em\hbox{\rm B}}}}
\def\IN{{\hbox{{\rm I}\kern-.2em\hbox{\rm N}}}}
\def\IC{{\ \hbox{{\rm I}\kern-.6em\hbox{\bf C}}}}

\def\IZ{{\hbox{{\rm Z}\kern-.4em\hbox{\rm Z}}}}
%
%Definition of arrows
%
\def\to{\rightarrow}

\def\onnearrow#1{\mathrel{\mathop{\nearrow}\limits^{#1}}}
\def\undernearrow#1{\mathrel{\mathop{\nearrow}\limits_{#1}}}
\def\onarrow#1{\mathrel{\mathop{\longrightarrow}\limits^{#1}}}

\def\A{{\cal A}}

\def\J{{\cal J}}
\def\bJ{{\overline{\cal J}}}
\def\O{{\cal O}}
\def\bO{{\overline{\cal O}}}
\def\Q{{Q}}
\def\bQ{{\overline{Q}}}
\def\T{{\cal T}}
\def\bT{{\overline{T}}}
\def\bz{{\bar{z}}}

\def\pd{{\partial}}
\def\bpd{{\bar{\partial}}}
\def\bY{{\overline{Y}}}
\def\bX{{\overline{X}}}
\def\bP{{\overline{P}}}
\def\bW{{\overline{W}}}
\def\res{{1\over 2 \pi i}}
\def\CINT{{\oint_{\cal C}}}
\def\C{{\cal C}}
\baselineskip 13pt
%%%%%%%%%%%%%%%%%%%%%%%%%%%%%%%%%%%%%%%%%%%%%%%%%%%%%%%%%%%%%%%%%%%%%
\pubnum{IASSNS-HEP-92/4 \cr
MIT-CTP-2057}
\date{January 1992}
\titlepage
\title{ALGEBRAIC STRUCTURES AND DIFFERENTIAL GEOMETRY \break
\break IN 2D STRING THEORY}
\author{Edward Witten\foot{Supported in part by NSF grant PHY91-06210.}
and Barton Zwiebach
\foot{Permanent address: Center for Theoretical Physics, MIT,
Cambridge, Mass. 02139. Supported in part by
D.O.E. contract DE-AC02-76ER03069 and NSF grant PHY91-06210.}}
\address{School of Natural Sciences\break
Institute for Advanced Study\break
Olden Lane\break
Princeton, NJ 08540}
\abstract{A careful treatment of closed string BRST cohomology shows
that there are more discrete states and associated symmetries in $D=2$
string theory than has been recognized hitherto.
The full structure, at the $SU(2)$ radius, has a natural description in terms
of abelian gauge theory on a certain three dimensional cone $Q$.
We describe precisely how symmetry currents are constructed from the
discrete states, explaining the role of the ``descent equations.''
In the uncompactified theory, we compute the action of the symmetries
on the tachyon field, and isolate the features that lead to nonlinear
terms in this action. The resulting symmetry structure is interpreted
in terms of a homotopy Lie algebra.}
\endpage

\chapter{Introduction}

Of the known soluble string theories in $D\leq 2$, the $D=2$ model
is in many
ways particularly intriguing.  It has a (two dimensional) space-time
interpretation in terms of the interactions of a massless scalar
field, somewhat misleadingly called the ``tachyon,''
and the space-time physics is realistic enough to include
black holes.

\REF\grossnewman{D. J. Gross, I. R. Klebanov, and M. J. Newman,
``The Two Point Correlation Function Of The One Dimensional Matrix
Model,'' Nucl. Phys. {\bf B350} (1991) 621; D. J. Gross and
I. R. Klebanov, Nucl. Phys. {\bf B359} (1991) 3.}
\REF\polyakov{A. M. Polyakov, ``Self-Tuning Fields And Resonant
Correlations In $2D$ Gravity,'' Mod. Phys. lett. {\bf A6} (1991) 635.}
\REF\lianzuck{B. Lian and G. Zuckerman, ``New Selection Rules And
Physical States in $2D$ Gravity,'' Phys. Lett. {\bf B254} (1991) 417;
{\bf B266} (1991) 21.}
\REF\others{S. Mukherji, S. Mukhi, and A. Sen,
Phys. Lett. {\bf B266} (1991) 337.}
\REF\bouwknegt{P. Bouwknegt, J. McCarthy, and K. Pilch, ``BRST Analysis
Of Physical States For $2D$ Gravity Coupled to $c\leq 1$ Matter,''
preprint CERN-TH 6162/91 (July, 1991), to appear in Comm. Math. Phys.}
\REF\itoh{K. Itoh and N. Ohta, ``BRST Cohomology And Physical
States in $2D$ Supergravity Coupled to $\widehat c\leq 1$ Matter,''
FERMILAB-PUB-91/228-T,OS-GE 20-91, Brown-HET-834 (September, 1991).}
\REF\witten{E. Witten, ``Ground Ring Of Two Dimensional String Theory,''
IASSNS-HEP-91/51, to appear in Nucl. Phys. B.}
\REF\kms{D. Kutasov, E. Martinec, and N. Seiberg, ``Ground Rings and their
Modules in 2D Gravity with $c \leq 1$ Matter,'' PUPT-1293, December 1991}
\REF\avanjevicki{J. Avan and A. Jevicki, Phys. Lett. {\bf B266}
(1991) 35.}
\REF\mandalwadia{S. R. Das, A. Dhar, G. Mandal, and S. R. Wadia,
IASSNS-HEP-91/72,79.}
\REF\mooreseiberg{G. Moore and N. Seiberg, ``From Loops To Fields In
Two Dimensional Gravity,'' Rutgers preprint RU-91-29 (1991).}
\REF\minicetal{D. Minic, J. Polchinski, and Zhu Yang,
``Translation-Invariant
Backgrounds in $1+1$ Dimensional String Theory,'' UTTG-16-91.}
\REF\klebpol{I. R. Klebanov and A. M. Polyakov, Mod. Phys. Lett. {\bf A6}
(1991) 3273}
In addition to the tachyon, the model also has ``discrete states''
which appeared
in the matrix model calculations of Gross, Klebanov,
and Newman [\grossnewman].
Polyakov proposed [\polyakov] that these states, which are what
survive in $D=2$ from the infinite tower of string states for $D>2$,
should be described by a sort of stringy topological field theory.
This suggestion was part of the motivation for subsequent efforts.
The discrete states originally considered were spin one currents,
but they have spin zero analogs found in the mathematical
analysis [\lianzuck--\itoh].
The spin zero (and ghost number zero) states
generate a ``ground ring'' [\witten]
which is characteristic of $D=2 $ string theory.  Simple considerations
involving the ground ring
explain many aspects of the free fermion description
that comes from the matrix model; this was argued in [\witten] and enlarged
upon (and extended to $D<2$) by Kutasov, Martinec, and Seiberg
[\kms].
In fact, at the $SU(2)$ radius, the ground ring is the ring of functions
on a certain three dimensional
cone $Q$; the three dimensions correspond in the matrix
model to the time, the matrix eigenvalue, and its canonical momentum.

Moreover, by combining the spin zero and spin one states,
one can construct an enormous unbroken symmetry group of the $D=2$
string theory.
At the $SU(2)$ radius, the group that arises is the group of volume
preserving diffeomorphisms of $Q$ (plus additional symmetries that
we will find in this paper).
In the uncompactified theory,
one gets essentially the $W_\infty$ symmetry of the matrix model
free fermions, described by several groups
[\avanjevicki--\minicetal].
For open strings, a $W_\infty$ current
algebra can be constructed, and the structure constants explicitly calculated,
using the original spin one discrete states
[\klebpol].

\REF\csft{See, for example, B. Zwiebach, Annals of Phys.
{\bf 186} (1988) 111}
\REF\nelson{P. Nelson, Phys. Rev. Lett. {\bf 62} (1989) 993;\hfill\break
H. S. La and P. Nelson, Nucl. Phys. B332 (1990) 83}
\REF\operatorformalism{L. Alvarez-Gaume, C. Gomez, G. Moore and
C. Vafa, Nucl. Phys. B303 (1988) 455;\hfill\break
C. Vafa, Phys. Lett. B190 (1987) 47}
\REF\distler{J. Distler and P. Nelson, Comm. Math. Phys. 138 (1991) 273}

In the present paper, we will analyze some aspects of this story
in somewhat more depth.  In \S2, we will look more closely at
the chiral BRST cohomology at the $SU(2)$ radius.  The chiral ground
ring is the ring of functions on a certain $x-y$ plane.  We will show
how certain peculiarities originally uncovered in mathematical
analysis of this cohomology can be described in terms of the differential
geometry of the $x-y$ plane.  In \S3, we combine left and right movers.
We show that this process is more subtle than usually supposed; the proper
analysis depends on the results of \S2 and on certain details
of closed string theory that are known in principle but do not usually
arise in practice [\csft--\distler ].
As a result we find
that (if the BRST cohomology is taken in the usual space of conformal fields)
there are more physical discrete states and symmetries at the $SU(2)$
radius than has been usually supposed.  All basic formulas
can be written in terms of the differential geometry of $Q$.  For instance
(overcoming a contradiction implicit in [\witten]),
the cubic couplings of discrete states at the $SU(2)$ point
can be generated by a Lagrangian
$$ L=\int \sigma \cdot F\wedge F,     \eqn\milmo$$
where $\sigma$ is a scalar and $F=dA$ is the field strength of an abelian
gauge field $A$.  The $SU(2)$ point is $\sigma=F=0$.

The enhanced symmetry of two dimensional
string theory at the $SU(2)$ radius\foot{Or any rational multiple of
that radius, where a similar structure arises.} is thus
somewhat analogous to the enhanced
symmetry of four dimensional general relativity, with Lagrangian
$$L={1\over 16\pi G}\int \epsilon_{abcd}e^a\wedge e^b\left(d\omega+\omega
\wedge \omega\right)^{cd}       \eqn\firsteq$$
at $e=\omega=0$.  This accounts for the theoretical significance of the
$SU(2)$ point and justifies our focussing on it in much of this paper.

\REF\ufwitten{E. Witten,  Nucl. Phys. {\bf B311} (1988/9) 46;
{\bf B323} (1989) 113.}
Of course, it is hard to make sense of general relativity expanded around
$e=\omega=0$.
(In three space-time dimensions one can do this
[\ufwitten], but this is related to the absence of local dynamics
in three dimensional general relativity.)
Two dimensional string theory seems to be our best example, at the moment,
of a model in which one can make some sense (at the $SU(2)$ radius)
of the analog of $e=\omega=0$, and one can also explicitly see
(in the uncompactified theory) a phase with local dynamics in the form
of tachyon scattering amplitudes.

It will be clear that our discussion of these matters is
preliminary and barely scratches the surface.
We will point out a few of the more obvious gaps at the end of \S3.2
and elsewhere.

In \S4, we explain how symmetry currents are constructed from
the discrete states -- using the ``descent equations.'' Of special
significance are the symmetry currents of ghost number zero which
are seen to arise from the BRST invariant states of ghost number one.
We also work out a number of explicit examples.

In \S5, we consider the uncompactified theory.  In particular, we analyze
how the symmetries associated with the discrete states act on the
tachyon field in the linear approximation.
In a suitable sense the tachyon field has spin one.
We also show that this result agrees with the prediction of the matrix
model description.

\REF\stasheff{J. D. Stasheff, Trans. Amer. Math. Soc. {\bf 108}
(1963) 293.}
\REF\getzler{E. Getzler and J. D. S. Jones, Ill. Jour. Math.
{\bf 34} (1990) 256.}
\REF\kontsevich{M. Kontsevich, ``Graph Cohomology,'' to appear.}

{}From the matrix model it is clear that the symmetries have the unusual
property of acting nonlinearly on the tachyon field.  In \S6,
we analyze in principle how this nonlinear action arises from the point
of view of conformal field theory, and illustrate the point by a simple
calculation.  We interpret the nonlinear terms in the Ward identities
in terms of homotopy Lie algebras [\stasheff--\kontsevich].
In \S7 we give some additional comments on our results.

The main points in the present work are probably that many of the structures
we find can be naturally described in terms of the differential geometry
of $Q$, and that many of them are similar to the structures arising
in BRST closed string field theory.

\chapter{The Chiral BRST Cohomology}

At the $SU(2)$ radius, with world-sheet cosmological constant zero,
the world-sheet Lagrangian of $D=2$ string theory is
$$L={1\over 8\pi}\int d^2x\sqrt h\left(h^{ij}\partial_iX\partial_jX
+h^{ij}\partial_i\phi\partial_j\phi\right)-{1\over 2\pi\sqrt 2}\int d^2x
\sqrt h \cdot \phi R^{(2)}. \eqn\secondo$$
($h$ is the world-sheet metric, $R^{(2)}$ is the Ricci scalar, and
$\phi$ is the Liouville field.)  At the $SU(2)$ radius, the allowed
values of the $X$ momentum are $p=n\sqrt 2$, $n\in \IZ / 2$.
Since the world-sheet cosmological constant $\mu$ is zero,
$X$ and $\phi$ are free, the left and right movers of the
theory are decoupled, and the BRST cohomology can be computed by first
studying the chiral problem, that is the right movers (or left movers)
only.  This has been worked out in detail [\lianzuck,\others,\bouwknegt];
we will recall the relevant points and work out some relevant consequences.
The reader may want to consult [\witten] for some background.

At spin zero and ghost number zero, BRST cohomology classes arise
precisely at discrete momenta $(p_X,p_\phi)=
(n,iu)\cdot \sqrt 2$, with $u=0,1/2,1,\dots$
and $n=u,u-1,\dots, -u$.
They are denoted as $\O_{u,n}$ and the first few are
$$\eqalign{
{\cal O}_{0,0} & = 1          \cr
x={\cal O}_{1/2,1/2} & = \left( cb+{i\over \sqrt 2}
(\partial X-i\partial\phi) \right)
    \cdot e^{i(X+i\phi)/\sqrt 2 } \cr
y={\cal O}_{1/2,-1/2} & = \left(cb-{i\over \sqrt 2}
(\partial X+i\partial\phi) \right)
    \cdot e^{-i(X-i\phi)/\sqrt 2 } .\cr     } \eqn\terzo$$
These states generate  under operator products the chiral ground
ring, which is simply the ring of polynomials in $x$ and $y$.

At spin zero and ghost number one, one has discrete states of the form
$$Y_{s,n}^{\pm}=c V_{s,n}\cdot e^{\sqrt 2 \phi \mp s\phi\sqrt 2} ,
\eqn\quarto$$
with $s=0,1/2,1,\dots $ and $n=s,s-1,\dots -s$.  Here $V_{s,n}$ is a primary
field constructed from $X$.  For the time being we are mainly interested
in the $Y_{s,n}^+$.  Note that for every ${\cal O}_{u,n}$, there is a
$Y^+_{u+1,n}$
with the same momentum; in addition there are the $Y_{s,\pm s}^+$
that do not have partners.  These are the ``discrete tachyons,'' that
is the tachyon modes that survive at the $SU(2)$ radius.

At these values of the momenta, the ${\cal O}$'s and $Y^+$'s actually
make up only half of the BRST cohomology.
This follows from the existence of the operator
$$ a=[Q,\phi]=c\partial \phi+\sqrt 2\partial c. \eqn\jolopo$$
$\phi$ is not a conformal field in the usual sense, but $a$ is.
Obviously, $a$ is BRST invariant, and in the usual space of conformal
fields it cannot be written as $[Q,\dots ]$.  $a$ is also a conformal
primary field.  In the present paper,
we will consider only the BRST cohomology in the usual space of conformal
fields, though this may be a restriction that should be
relaxed.

This being so, we can form new families of BRST invariant vertex
operators.  At ghost number one we can extract a new operator
which we will call $a{\cal O}_{u,n}$ from
the operator product of $a$ and ${\cal O}_{u,n}$.
To be precise,
$$a{\cal O}_{u,n}(0)
={1\over 2\pi i}\oint {dz\over z}a(z)\cdot {\cal O}_{u,n}(0).
\eqn\offo$$
Obviously, $a{\cal O}_{u,n}$ has the same momenta as ${\cal O}_{u,n}$.
Similarly by multiplying $a$ and $Y_{s,n}^+$ we make a new family
of spin zero,
ghost number two operators, which we will call $aY_{s,n}^+$.
(The case $s=0$ is exceptional here and in many later statements.)
The operators made this way are
all nonzero and independent of the
old ones.  This follows from the treatment of the BRST cohomology
in [\bouwknegt],
and we will in any case soon exhibit an inverse to multiplication
by $a$.  These states and their conjugates with opposite Liouville dressing
are known to exhaust the BRST cohomology.

The BRST cohomology at typical discrete momenta can be arranged in
a diamond ($n \leq u$)
$$\matrix{ &  & a{\cal O}_{u,n} &  &  \cr
 & \onnearrow{a\ \ \ \ } &  & \searrow &  \cr
{\cal O}_{u,n} &  &  &  & aY^+_{u+1,n} \cr
 & \searrow &  & \undernearrow{\ \ a} &  \cr
 &  & Y^+_{u+1,n} &  &  \cr\cr }$$
There are four states,
of ghost numbers $(0,1,1,2)$.  At the special tachyon momenta
$(\pm s, i(s-1))\cdot \sqrt 2$, there are only two states, namely
$Y_{s,\pm s}^+$ and $aY_{s,\pm s}^+$:
$$
\matrix{ &  & Y^+_{s,\pm s} & \onarrow{a} & aY^+_{s,\pm s} \cr} .$$
The latter turns out to be
(for $s\not= 0$)
$$aY_{s,\pm s}^+\sim c\partial c V_{s,\pm s}\cdot e^{\sqrt 2 \phi (1-s)}.
       \eqn\holipo$$

\subsection{The ``Minus'' States}

Now let us briefly discuss the ``minus'' states, that is the states
whose Liouville dependence is $e^{\sqrt 2\phi(1+s)}$ with $s\geq 0$.
The two point function on the sphere gives a pairing
$$\langle {\cal V}_i^-{\cal W}_j^+\rangle     \eqn\hoffo$$
between the minus and plus states.  This pairing is nondegenerate
and has ghost
number three (from the three chiral ghost zero modes
on the sphere) and Liouville momentum $(-2i\sqrt 2)$ (from the Liouville
background charge).  So the ``dual'' of a ``plus'' state
with ghost number $j$ and Liouville factor $e^{\sqrt 2\phi(1-s)}$
is a state with ghost number $3-j$ and Liouville factor
$e^{\sqrt 2\phi(1+s)}$.  As the
ghost numbers of the ``plus'' BRST cohomology
range from 0 to 2, those of the ``minus'' states range from 1 to 3.
The ``minus'' states form patterns dual to those of the ``plus'' states
$$
\matrix{ &  & aY^-_{u+1,n} &  &  \cr
 & \onnearrow{a\ \ \ \ } &  & \searrow &  \cr
Y^-_{u+1,n} &  &  &  & a P_{u,n} \cr
& \searrow &  & \undernearrow{\ \ a} &  \cr
 &  & P_{u,n}  &  &  \cr\cr
 &  Y^-_{s,\pm s} & \onarrow{a} & aY^-_{s,\pm s} & \cr}$$
\vskip .1in
The minus states
of most immediate interest to us will be those of ghost number 1.
These are simply the $Y_{s,n}^-$ (which are the duals of
the $aY_{s,n}^+$).
As explained in [\witten,\S2.4], these states
transform under area preserving diffeomorphisms
as derivatives of a delta
function supported at the origin of the $x-y$ plane.

\section{Interpretation of the Discrete States}

The discrete states can be given the following interpretation, which
may seem a nicety to begin with but will prove to be important.
The states at ghost number 0 are just the polynomial functions on the
$x-y$ plane, as explained in [\witten]. (The operator
$\O_{u,n}$ corresponds to the function $x^{u+n}y^{u-n}$).
The $x-y$ plane is endowed with the area form
$$\omega = dx\wedge dy. \eqn\yorox$$
The $Y_{s,n}^+$ with $s>0$ correspond, in a sense explained
in [\witten], to polynomial vector fields on the $x-y$ plane that
generate area-preserving diffeomorphisms.
(If ${\cal O}_{u,n}$ corresponds to a function $f$, then $Y^+_{u,n}$
corresponds to the area preserving
vector field $V^i=\omega^{ij}\partial_jf$.)
Let $V$ be any
vector field on the $x-y$ plane, and ${\cal L}_V$ the corresponding Lie
derivative.  Then ${\cal L}_V(\omega)=f\cdot  \omega$ where
$f=\partial_iV^i$
is a function.  $V$ is uniquely determined by $f$ modulo addition
of an area-preserving vector field; and every $f$ can arise.
Thus, once the area preserving vector fields
have been identified with the $Y$'s, the additional vector fields on the
$x-y$ plane have the quantum
numbers of the functions.  Since $a$ has momentum zero, the states
$a{\cal O}_{u,n}$ have the same momenta as the ${\cal O}$'s and are thus
in natural correspondence with the functions on the $x-y$ plane.
As regards the momentum quantum numbers,
they  are the missing operators that we need to make up
{\it all} the (polynomial) vector fields on the $x-y$ plane.
So, for counting states, the ghost number one operators
$Y_{s,n}^+$ and $a{\cal O}_{u,n}$ make up the arbitrary vector fields
on the plane.

At ghost number two, we have the $aY_{s,n}^+$, which have the same
momenta as the ${\cal O}_{s,n}$, except for a shift in the Liouville
momentum by $-i\sqrt 2$.  Since $\partial/\partial x_i$ has the opposite
quantum numbers of $x^i$, $(0,-i\sqrt 2)$
are precisely the $(p_X,p_\phi)$ values of the bivector
$${\partial\over\partial x}\wedge {\partial\over\partial y}. \eqn\melba$$
So, as the ${\cal O}_{u,n}$'s have the quantum numbers of polynomial
functions, the
$aY_{s,n}^+$'s have the quantum numbers of the polynomial bivector fields on
the
plane.

These results, which we have obtained piecemeal, can be presented
in the following unified way.  Let $T$ be the tangent bundle of the $x-y$
plane.  Let $\wedge^iT$, $0\leq i\leq 2$ be its $i^{th}$ exterior power.
Then the discrete states of ghost number $i$, for $i=0,1,2$, transform
as the polynomial sections of $\wedge^iT$.  In other words, one has
functions, vector fields, and bivectors for $i=0,1,2$.

The BRST cohomology has a natural ring structure -- induced from the
operator products -- even if we do not restrict to ghost number zero.
The multiplication law going from ghost number $i$ times ghost number
$j$ to ghost number $i+j$ is just the natural ``wedge product''
$\wedge^iT\times \wedge^jT\to \wedge^{i+j}T$, for $0\leq i,j\leq 2$.
This can be verified using arguments similar to the ones given for
$i=j=0$ in [\witten].

\subsection{The Cosmological Constant}

There is, in addition, one more operator that
does not fit into this framework,
namely the cosmological constant operator $\iota=Y_{0,0}^+=Y_{0,0}^-$.
This is of ghost number one, and so should correspond to a symmetry
charge (as we will show in \S4 ).
However, as explained in [\witten], this operator is central.
This is why we give it the name $\iota$, which is meant to be suggestive
of the fact that the central operator in ordinary quantum mechanics is
customarily called $i$.

One cannot make an operator of ghost number two by acting on $\iota$
with $a$, since, as is easily verified, $a\cdot \iota = 0$.  Nevertheless,
a corresponding operator of ghost number 2, namely $c\partial c e^{\sqrt
2\phi}$, does exist.
This is the operator which, in the construction above, corresponds to the
constant bivector
$${\partial\over\partial x}\wedge {\partial\over\partial y}. \eqn\ungo$$
We will call this operator $\widetilde \iota$; occasionally, to simplify
various statements, we may somewhat inconsistently call it
$a\cdot Y_{0,0}^+$.

\subsection{The Dual Version}

There is a dual version of this, making use of the area form $\omega$,
which will prove to be essential.  Contraction with $\omega$ is a natural
operation mapping $i$-vectors, for $i=0,1,2$, to differential forms
of degree $2-i$.  Thus, one maps the 0-vector or function $f$ to the
two form $f\omega$; the vector field $V^i\partial_i$ to the one form
$V^i\omega_{ij}dx^j$; and the bivector
$$ p^{ij}{\partial\over\partial x^i}\wedge{\partial\over\partial x^j}
\eqn\bive$$
to the function $p^{ij}\omega_{ij}/2$.
So in particular,  the constant bivector \ungo\ is mapped to the function 1.

Hence, we could think of the discrete states of ghost number $i$
as $2-i$ forms.  This point of view seems unnatural from what we have
said so far, because it obscures the ring structure.
But facts that will presently appear shed a different light.

\section{Operators And Currents}

Under some circumstances, from a BRST cohomology class of ghost number $g$,
one can construct
a conserved current of ghost number $g-1$.

This is done as follows.  Consider a BRST cohomology class represented
by a highest weight BRST invariant state $|\psi\rangle$
of $L_0=0$.
Set $|\alpha\rangle =b_{-1}|\psi\rangle$
(and assume $|\alpha\rangle\not=0$).
Then $L_0|\alpha\rangle=|\alpha\rangle$, so if $|\alpha\rangle$
is a highest
weight state, the highest weight is one, and the operator corresponding
to $|\alpha\rangle$ is a current.  Moreover,
$$Q|\alpha\rangle =Qb_{-1}|\psi\rangle = L_{-1}|\psi\rangle.  \eqn\opfo$$
Hence, although the current corresponding to $|\alpha\rangle$ is not BRST
invariant, its BRST commutator is a total derivative.
To see whether $|\alpha\rangle$ is a highest weight state at least
modulo BRST commutators, note that
$$L_n|\alpha\rangle = L_nb_{-1}|\psi\rangle =b_{n-1}|\psi\rangle. \eqn\offor$$
These states may not vanish,  but we do at least have $Qb_{n-1}|\psi\rangle
=L_{n-1}|\psi\rangle =0,\,\,\,\,n>0$, as $|\psi\rangle$ is BRST invariant
and highest weight.  Therefore, if the BRST invariant states $b_{n-1}|\psi
\rangle$ are BRST commutators, then $|\alpha\rangle$ is of highest weight,
at least modulo BRST commutators.

For $n>1$, this is automatically so, as
$$b_{n-1}|\psi\rangle=-{1\over n-1}Q b_0 b_{n-1}|\psi\rangle.  \eqn\ippo$$
The crucial case is therefore $n=1$.  $|\alpha\rangle$ is of highest weight
(modulo BRST commutators) if $b_0|\psi\rangle =Q|\dots\rangle$.

Thus, the linear transformation $|\psi\rangle \to b_0 |\psi\rangle$
of the BRST cohomology plays a crucial role. The cohomology classes that
give rise to currents, in the chiral theory, are those that are in the
kernel of $b_0$.

\subsection{Status Of The Discrete States}

Now let us apply this to the discrete states of the $D=2$ model.

First of all, the states corresponding to the operators
${\cal O}_{u,n}$ would be mapped by $b_0$ to ghost
number $-1$, where the BRST cohomology vanishes.  So they
are annihilated by $b_0$, at least at the level of cohomology.
In the appendix, we show that this is true exactly.
These states therefore give rise to highest weight
currents of ghost number $-1$.

The $Y_{s,n}^+$ are likewise annihilated by $b_0$.  To see this, note
that at the level of operators, the operation of multiplying by $b_0$
can be written
$$V(0)\to b_0 \cdot V(0) \equiv
{1\over 2\pi i}\oint dz\,\, z \,\,b(z)\cdot V(0). \eqn\riffo$$
This removes from the states a factor of $\partial c$, wherever this
is present, and otherwise gives zero.
The explicit form of the wave functions of the $Y$'s shows
that there are no terms in $\partial c$, and hence
$b_0 \cdot Y_{s,n}^+ =0$.
The $Y_{s,n}^+$ thus give rise to currents;
these generate area preserving diffeomorphisms (of the $x-y$ plane),
as shown in [\witten,\klebpol].

Multiplication by $b_0$ is not a trivial operation, however.
One immediately sees that, with $a$ as defined in \jolopo,
$$b_0\cdot a = \sqrt 2.      \eqn\forgo$$
This strongly suggests that if a discrete state $X$
is annihilated by $b_0$, then
$a\cdot X$ is not.  This is actually
true (except for the cosmological constant operator),
as follows from the analysis of the absolute and relative BRST
cohomology in [\lianzuck,\others].  The relevant calculation can
be conveniently done on a cylinder with angular variable $\theta$.
We have
$$b_0={1\over 2\pi}\oint d\theta \,\,\,\,\,b(\theta).     \eqn\orgo$$
Multiplication by $a$ can be represented by the operator
$$ \widetilde a ={1\over 2\pi}\oint d\theta\,\,\,\,
a(\theta)={1\over 2\pi}\oint d\theta
\left(c\partial\phi+\sqrt 2\partial c\right). \eqn\rgo$$
Hence
$$[b_0,\widetilde a] ={1\over 2\pi}\oint d\theta \,\,\,\,\,
                                  \partial\phi. \eqn\ggo$$
This is the operator that measures the Liouville momentum.
Hence if $b_0\cdot {\cal O}=0$, and ${\cal O}$ corresponds to a Fock space
state $|\psi\rangle$ of Liouville momentum $p_\phi(\psi)$, then
$b_0\cdot(a{\cal O})$ corresponds to the Fock space state
$p_\phi |\psi\rangle$.
Bearing in mind
the shift in Liouville momentum by $-i\sqrt 2$ in going from operators
to states defined on the cylinder, $p_\phi=0$ only for the cosmological
constant operator $\iota$; and so that is the only case for which $b_0$
annihilates $a{\cal O}$.  Actually,
$a\iota=0$, and that is why $b_0$ annihilates $a\iota$;
$b_0$ does not annihilate the operator $\widetilde \iota=
c\partial c e^{\sqrt 2\phi}$.

Notice that the current $\partial \phi$ that appeared in the above
computation is not a primary field (as $a$ is not annihilated by $b_0$).
The above computation was formulated from the start with an angular
variable $\theta$ on the cylinder, and it was not guaranteed that the
operators arising in the computation would be primary.

\section{The $b_0$ Operator as an Exterior Derivative}

What sort of operator is $b_0$?  For instance, does it
commute with the action of the ground ring?

Let ${\cal O}$ be a spin zero, ghost number zero operator of the
ground ring.  We can represent ${\cal O}$ by the operator
$$\widehat {\cal O}= \CINT {dw \over 2\pi i}\,\,{{\cal O} \over w}.
\eqn\turof$$
The commutator of this with $b_0$ is
$\left[b_0,\widehat{\cal O}\right]$. This object is actually BRST
invariant: $\{ Q , \left[b_0,\widehat{\cal O}\right]\}$ vanishes
because $\widehat{\cal O}$ is BRST invariant and, being of dimension
zero, it commutes with $L_0 = \{ Q , b_0 \}$.
Our object $\left[b_0,\widehat{\cal O}\right]$ is actually a charge,
that is the integral of a spin one operator,
since in computing this commutator a contour integral remains.
There are non-trivial BRST invariant charges of ghost number $-1$
(as we have seen in discussing the existence of currents), and there
is no reason to expect $\left[b_0,\widehat{\cal O}\right]=0$.
Indeed, this is not true,
as one readily sees in simple examples (e.g., ${\cal O}=x$).

The fact that the $b_0=0$ condition does not commute with the ${\cal O}$'s
means that the currents -- which correspond to the part of the cohomology
annihilated by $b_0$ -- do not form a module for the ground ring.
For essentially this reason,  when we combine left and right movers,
the discrete moduli of the theory will not form a ground ring module.
One might think that the utility of the ground ring would thereby be lost.
What saves the day is that $b_0$ obeys the following condition.
Let ${\cal O}$ and ${\cal O}'$ be any two (spin zero, ghost number zero)
ground ring states.
Then
$$[ [b_0,\widehat{\cal O}],\widehat{\cal O}']=0
\eqn\huffo$$
modulo $\{Q,\dots\}$.
The point is that
$[[b_0,\widehat{\cal O}],\widehat{\cal O}'(w)]$ would be a BRST
invariant {\it local} operator of ghost number $-1$ and $L_0=0$; but the BRST
cohomology with those quantum numbers is trivial.

What is the significance of \huffo?  We recall that the ${\cal O}$'s
have the interpretation of multiplication by functions on the $x-y$ plane.
In general, if an operator $X$ has the property that $[[X,f],g]=0$
for any two functions $f$ and $g$, then $X$ is a first order differential
operator.  Hence, \huffo\ means that $b_0$ can be interpreted as a first
order differential operator on the $x-y$ plane.

$b_0$ maps states of ghost number $q$ to states of ghost
number $q-1$, for $q=1,2$, so (from our analysis of the quantum numbers)
it maps $\wedge^qT$ to $\wedge^{q-1}T$.  Dually, if we use the area form
to identify $\wedge^qT$ with the space $\Omega^{2-q}$ of $2-q$ forms,
then $b_0$ maps $\Omega^i$ to $\Omega^{i+1}$.  Furthermore, $b_0{}^2=0$.
In addition, $b_0$ must commute with symmetries of the theory, which
are a central extension of the polynomial area preserving diffeomorphisms
of the plane.  The central extension is described by the formula
$$\left[{\partial\over \partial x},{\partial\over\partial y}\right]
=\iota \eqn\duco$$
of [\witten, eqn. 2.23].  Ignoring the central extension (setting $\iota$ to
0, so to speak), there
is only one first order operator on the $x-y$ plane that commutes with
area-preserving diffeomorphisms; this is the exterior
derivative $d$.  The central extension necessitates the following
slight modification: if $\alpha$ is a $j$ form, then
$$b_0\alpha=d\alpha +\delta_{j=0}\alpha(0) \iota.      \eqn\ilpo$$
(That is, the $\iota$ dependent correction vanishes unless $\alpha$ is a zero
form, in which case $\alpha(0)$ is the function $\alpha$ evaluated at
$x=y=0$.)

Let us check that this identification gives the right answer for the
kernel of $b_0$.  Acting on zero forms, the exterior derivative annihilates
only the constants, which however are not annihilated by the $\iota$
dependent part of $\ilpo$.  In our identification of discrete states with
differential forms on the plane, zero forms correspond to ghost number
two, and we have indeed seen that no states of ghost number two are
annihilated by $b_0$.  As for one forms, as the $x-y$ plane is contractible,
a closed one form $\lambda$ on the $x-y$ plane is $\lambda=df$ for some
$f$.  In the relation between $\Omega^1$ and $\wedge^1T=T$, $\lambda$
corresponds to the vector field
$$V^i{\partial\over\partial x^i}=-\partial_if\cdot \omega^{ij}\cdot
{\partial\over\partial x^j}. \eqn\gurgo$$
This is an area preserving vector field derived from the Hamiltonian
function $f$.  Thus, the currents, which are associated with the kernel
of $b_0$, correspond to those vector fields on the plane which are
area preserving.  Indeed, the currents
derived from the $Y_{s,n}^+$ were seen in [\witten,\klebpol] to generate
the group of area preserving transformations (of the $x-y$ plane).

We summarize in the table below the interpretation we have derived
in this section for the states
in the BRST chiral cohomology.
\vskip .1in
$$\hbox{\vbox{\offinterlineskip
\def\strut{\hbox{\vrule height 15pt depth 10pt width 0pt}}
\hrule
\halign{
\strut\vrule#\tabskip 0.1in&
\hfil$#$\hfil &
\vrule#&
\hfil$#$\hfil &
\vrule#&
\hfil$#$\hfil &
\vrule#&
\hfil$#$\hfil &
\vrule#\tabskip 0.0in\cr
& && \hbox{Ghost \#} && \hbox{Interpretation} && \matrix{\hbox{Dual}\cr
\hbox{Interpretation}\cr} & \cr\noalign{\hrule}
& {\cal O} && 0 && \hbox{functions} && \hbox{two-forms} & \cr\noalign{\hrule}
& Y^+ && 1 && \matrix{\hbox{area preserving}\cr\hbox{vector fields}\cr} &&
\hbox{closed one-forms} & \cr\noalign{\hrule}
& a{\cal O} && 1 && \matrix{\hbox{area non-preserving}\cr\hbox{vector
fields}\cr} && \matrix{\hbox{\ \ one-forms that\ \ }\cr\hbox{are not closed}
\cr} &
\cr\noalign{\hrule}
& aY^+ && 2 && \hbox{bivectors} && \hbox{zero-forms} &
\cr\noalign{\hrule}}}}$$
\medskip

\chapter{The Closed String Cohomology}

In this section, we will analyze the BRST cohomology of the closed
string in $D=2$, beginning with the $SU(2)$ point.  It is here that we will
enjoy the payoff from our work in the last section.

If $|\psi_L\rangle$ and $|\psi_R\rangle$ are BRST invariant states
in the left and right moving Fock spaces, with the same value of the
Liouville momentum $p_\phi$, then the tensor product $|\Psi\rangle
=|\psi_L\rangle\otimes |\psi_R\rangle$ is certainly a BRST invariant
state of the closed string.  Moreover, all closed string states arise
this way.\foot{Without the $b_0-\overline b_0$ condition that we will introduce
presently, this is trivial.  We will prove at the end of this section that
it is true even when this condition is imposed.}

The closed string ground ring, at ghost number zero, is therefore easy to
construct.  Let us recall how this is done [\witten].
The left and right moving ghost number zero states
are linear combinations of
$x^ny^m$ and $x'^{n'}y'^{m'}$, with $x,y$ and $x',y'$ being the
generators of the left and right moving chiral ground rings.
The equality of Liouville momenta is the condition $n+m=n'+m'$.
Products $x^ny^nx'^{n'}y'^{m'}$ with $n+m=n'+m'$ are monomials in
$a_1=xx'$, $a_2=yy'$, $a_3=xy'$, and $a_4=yx'$.  The $a_i$
obey one relation $a_1a_2-a_3a_4=0$. This equation defines a quadric
cone $Q$ in the $a_i$ space.  The closed string ground ring is the ring
of polynomials in the $a_i$, or in other words the ring of polynomial
functions on $Q$.

It is useful to introduce a four dimensional space $W$ with coordinates
$x,y,x',y'$.  On $W$ there is a one parameter group action
$x,y,x',y'\rightarrow tx,ty,t^{-1}x',t^{-1}y'$.  We will call this group
$H$.  $H$ is generated by the vector field
$$ S=x{\partial\over\partial x}+y{\partial\over\partial y}
  -x'{\partial\over\partial x'}-y'{\partial\over\partial y'}\eqn\offo$$
which measures the difference between the left and right moving Liouville
momenta.
$Q$ is just the quotient $W/H$.  Many constructions below are most
conveniently described in terms of $H$ invariant objects on the four
dimensional
space $W$, rather than explicitly working out the reduction to $Q$.

\subsection{The $b_0-\overline b_0$ Condition}

The construction of the ground ring is so trivial that one may well wonder
if there is anything non-trivial to be done in combining the left and right
movers.

There is; this becomes clear when one
considers the discrete moduli of the closed string.
Moduli are usually constructed from
primary fields of highest weights $(1,1)$ and left and right moving
ghost numbers $(0,0)$.  To construct these by multiplying left and right
movers, one needs a left moving current $Y$ and a right moving current
$\overline Y$, each of spin one and ghost number zero.  As we have
seen in \S2.2, these will arise as $|Y\rangle=b_{-1}|Z\rangle$
 and $|\overline Y\rangle=\overline
b_{-1}|\overline Z\rangle$, where $|Z\rangle$
and $|\overline Z\rangle$ have spin zero and ghost
number one and are annihilated by $b_0$ and $\overline b_0$, respectively.
The moduli that one gets this way correspond thus to
$|{\cal Y}\rangle=b_{-1}\overline b_{-1}|{\cal Z}\rangle$, where
$|{\cal Z}\rangle=|Z\rangle\otimes|\overline Z\rangle$.
Here $|{\cal Z}\rangle$ corresponds to an operator of spin $(0,0)$ and ghost
number $(1,1)$.
Moreover,
$$b_0|{\cal Z}\rangle=\overline b_0|{\cal Z}\rangle = 0 . \eqn\hipo$$
The discrete moduli that have been discussed
hitherto are the ones constructed as just explained.

That this cannot be the whole story can be seen in a number of ways.
First of all, in explicitly acting on moduli of the above type with
symmetry generators,
new operators appear.  This computation will be presented in \S5.3 below.
Another reason that new operators are needed comes from consideration
of the $\beta$ function; see \S3.2.
The reason for the problem is that although the construction of
closed string moduli recalled in the last paragraph
is the most familiar one, the general rules of closed string theory
permit a more general construction.
Closed string moduli are always derived from BRST invariant operators
${\cal Z}$ of ghost number 2 and spin zero.  However, the left
and right moving ghost numbers need not both be 1 in general.
Moreover, condition \hipo\ is stronger than necessary; the essential
condition is only
$$ \left( b_0-\overline b_0\right)\,\ket{\cal Z} = 0 . \eqn\ripo$$
That these are the only really necessary conditions is known in the
operator formalism [\nelson--\distler] (where it arises due to the
absence of a global section on ${\cal P}_{g,n}$) and
is actually obvious in closed string field theory [\csft]
(where this restriction is necessary to write a kinetic term).
In many situations,
nothing is lost (and life is easier) if one restricts to
${\cal Z}$'s that  obey \hipo\ and have ghost numbers $(1,1)$.
But in $D=2$, the story is completely different.

\section{Closed String Moduli}

We have seen that the left moving discrete states of ghost number
$i$ can be identified with the $2-i$ forms on the $x-y$ plane.
Similarly the right moving discrete states of ghost number
$j$ are the $2-j$ forms on the $x'-y'$ plane.

The product of the $x-y$ plane and the $x'-y'$ plane is the four
manifold $W$ introduced above.  By taking the tensor product
of arbitrary left and right moving discrete states, one obtains
the (polynomial) differential forms on $W$.
The degree of the differential form is $(2-i)+(2-j)=4-i-j$, where
$i$ and $j$ are the left and right moving ghost numbers.
If we denote the space of $n$ forms on $W$ as $\Omega^n$, and
the space of forms of type $i,j$ (with $i$ indices tangent to the
$x-y$ plane and $j$ to the $x'-y'$ plane) as $\Omega^{i,j}$, then
$$\Omega^n=\oplus_{i=0}^n\Omega^{i,n-i}. \eqn\toof$$
This just says that the total ghost number
is the sum of the independent left and right moving contributions.

In tensoring together left and right moving states, we should restrict
to states of equal left and right moving Liouville momenta.
This just means taking the $H$ invariant
states.  If, therefore, we do not worry about the
$b_0-\overline b_0$ condition, then the discrete states of the closed
string are just the $H$ invariant differential forms on $W$.

What about $b_0-\overline b_0$?  We have identified $b_0$ as the
exterior derivative on the $x-y$ plane -- let us call it $d_L$.
(For simplicity, we will ignore the central extension.)
Likewise $\overline b_0$ is the exterior derivative on the $x'-y'$
plane, say $d_R$.  Hence $b_0-\overline b_0=d_L-d_R$.  By conjugating
with $(-1)^{F_R}$ (the operator that multiplies $j$ forms on the
$x'-y'$ plane by $(-1)^j$), we can transform this into $d=d_L+d_R$,
the standard exterior derivative on $W$.   Note that
$$0=d_L{}^2=
d_R{}^2=\{d_L,d_R\}.\eqn\mollo$$

As we have discussed, closed string moduli correspond to
discrete states ${\cal Z}$
of ghost number two that are annihilated by $b_0-\overline b_0$.
These are in other words, in view of what we have just said, the
closed two forms on $W$ which are also $H$ invariant.

A closed two form $F$ is naturally regarded as the curvature of an
abelian gauge field.  The simple topology of $W$ or $Q$ does not support
Dirac monopoles, so
we can write simply $F=dA$, with $A$ being the vector potential.
We have found that the closed string moduli
(in the space of conformal fields) can be represented by an $H$ invariant
abelian gauge field on $W$.  Imposing the $H$ invariance means carrying
out a dimensional reduction from $W$ to the quotient space $Q=W/H$.
Under this process, $A$ reduces to an abelian gauge field and a scalar
on $Q$.  Many formulas are simpler if written in terms of $H$ invariant
objects on $W$, and we need not always carry out the dimensional reduction
explicitly.

Let us summarize the closed string states that obey the $b_0-\overline
b_0$ condition. For convenience we split the states according to
the type of Liouville dressing. The ``plus'' states are
$$\eqalign{
G &=0: \quad \O_{u,n} \bO_{u,n'}
\cr
G &=1: \quad Y^+_{s,n} \bO_{s-1,n'},\quad
\O_{s-1,n'}\bY^+_{s,n},\quad
(a+\bar{a})\cdot (\O_{u,n}\bO_{u,n'}),
\cr
G &=2: \quad Y^+_{s,n}\bY^+_{s,n'}, \quad
(a+\bar{a})\cdot (Y^+_{s,n}\bO_{s-1,n'}),\quad
(a+\bar{a})\cdot (\O_{s-1,n'}\bY^+_{s,n}),
\cr
G &=3: \quad (a+\bar{a})\cdot (Y^+_{s,n}\bY^+_{s,n'}),
\cr}\eqn\srpluscoh$$
and the ``minus'' states are
$$\eqalign{
G &=2: \quad Y^-_{s,n}\bY^-_{s,n'},
\cr
G &=3: \quad Y^-_{s,n} \bP_{s-1,n'},\quad
P_{s-1,n'}\bY^-_{s,n},\quad
(a+\bar{a})\cdot (Y^-_{s,n}\bY^-_{s,n'}),
\cr
G &=4: \quad P_{s,n} \bP_{s,n'},\quad
(a+\bar{a})\cdot (Y^-_{s,n}\bP_{s-1,n'}),\quad
(a+\bar{a})\cdot (P_{s-1,n'}\bY^-_{s,n}),
\cr
G &=5: \quad (a+\bar{a})\cdot (P_{s,n}\bP_{s,n'}).
\cr}
\eqn\srminuscoh$$

\section{Comparison To The Usual Operators}

In [\witten,\S2.6], it is pointed out
that the closed string moduli coming from conventional $(1,1)$ vertex
operators of ghost number zero (which were the only ones considered
there) correspond
to functions on $Q$.  Let us see how these fit in.

If we are given the function $\phi$ on $Q$, which we will think of as
an $H$ invariant function on $W$, we can make the two form
$$F=d_Ld_R\phi.         \eqn\offo$$
Explicitly
$$F_{ij'}={\partial^2\phi\over\partial x^i\partial x'^{j'}},
{}~~~~~F_{ij}=F_{i'j'}=0. \eqn\noffo$$
Thus, $F$ is a differential form of type $(1,1)$, and obviously
(using \mollo)
$d_LF= d_RF=0$.  Thus, the operator corresponding to $F$
has left and right moving ghost numbers $(1,1)$, and is annihilated
by both $b_0$ and $\overline b_0$.  These are then the conventional
$(1,1)$ moduli, the ones considered in previous discussions of
the discrete states.
Thus we have explained the embedding of the functions on $Q$ -- considered
in [\witten] -- in the closed $F$ invariant two forms on $W$, which
is the space of closed string moduli in the sense we are considering
here.

To first order the states we have been discussing are all closed
string moduli.  In quadratic order, one meets a variety of effects
including the quadratic part of the beta function.  As was noted
in [\witten, eqn. (2.47)], the quadratic part of the $\beta$ function
for the $\phi$ field is
$$\beta = \epsilon^{ij}\epsilon^{i'j'}{\partial^2\phi\over\partial x^i
\partial x'^{i'}}{\partial^2\phi\over\partial x^j\partial x'^{j'}}.
\eqn\turox$$
There is a puzzle here.  This formula is not invariant under volume
preserving diffeomorphisms of $Q$, which were claimed in [\witten]
to be a symmetry of the theory.  The puzzle is one reason that the
extra states that we have  been describing here are necessary.
With the formula \noffo\ for the map from the function $\phi$ to the
two form $F$, we see that in terms of $F$, the quadratic beta function
can be written
$$\beta = F\wedge F.             \eqn\urpo$$
This formula is invariant, as it should be, under volume preserving
diffeomorphisms; in fact it is invariant under arbitrary diffeomorphisms
of $W$ (that commute with $H$ if one restricts to the $H$ invariant subspace).

\subsection{The Minus Operators}

We now want to write down a Lagrangian from which \urpo\ can be derived.
To this aim, we must
consider the ``minus'' operators, the ones with Liouville
dressing $e^{\sqrt 2 \phi (1+s)}$, $s\geq 0$.  Here no new moduli
appear, compared to previous analyses, for the following reason.
As we have noted at the beginning of \S2, the left or right moving minus
operato
   rs
have ghost numbers in the range between 1 and 3.
Moreover, the states of ghost number 1 in the chiral theory are annihilated
by $b_0$ (or $\overline b_0$).
In combining such
left and right moving modes, the total ghost number
can be 2 only if the left and right moving pieces are $(1,1)$.
The modes made this way are annihilated by $b_0$ and $\overline b_0$
separately and are thus modes of the type considered in the past.

As the left and right moving zero modes can be interpreted as
``functions'' (or really distributions)
on the $x-y$ or $x'-y'$ plane with delta function
support at the origin, when one combines left and right movers
the ``minus'' moduli make up an $H$ invariant ``function'' on $W$ with support
at the origin (or a function on $Q$ with delta function support at the
apex).  Let us call this function $\sigma$.

\subsection{The Lagrangian}

The Lagrangian from which
\urpo\ is to be derived should involve both $\sigma$ and $F$, and is obviously
$$L=\int \sigma\cdot F\wedge F     . \eqn\rpo$$
The $\sigma$ equation of motion is the $F\wedge F$ part of the beta
function; the $A$ equation of motion gives terms in the beta function
bilinear in $\sigma $ and $F$. (We have verified these terms in part.)

The Lagrangian \rpo\ is oddly reminiscent of Lagrangians sometimes studied
in works on topological field theory.  It very much looks like the Lagrangian
of a theory that does {\it not} have local dynamics.  In many ways
the most remarkable and mysterious aspect of the theory is that the same
model that at the $SU(2)$ point gives rise to \rpo\ also gives rise
in the decompactified theory to the local dynamics of the ``tachyon.''

The absence of a quadratic term in \rpo\ is unsettling but, given the
framework for the computation, inevitable: \rpo\ is by definition a
cubic coupling of modes that are moduli in lowest order and hence appear
in no quadratic coupling.

Several caveats should be stated here:

(1) If we take the analysis at the $SU(2)$ point literally, the
moduli contained in $F$ correspond to forms with a polynomial
dependence on the coordinates $a_i$ of $Q$, while
$\sigma$ (which is constructed from states with the dual Liouville dressing)
is a ``function'' (or really a distribution) supported at $a_i=0$.
It is, however, very tempting to believe that with better understanding,
perhaps after making some generic perturbation of the model,
$F$ and $\sigma$ should be permitted to be general fields on
$Q$, rather than polynomials on the one hand and derivatives of a delta
function on the other.

(2) The Lagrangian \rpo\ correctly describes the cubic couplings of
discrete states at the $SU(2)$ point.  Whether it correctly describes
the restrictions needed to maintain conformal invariance under
departure from the $SU(2)$ point is less clear.  Once one starts to
depart from the $SU(2)$ point, it is not clear that the BRST cohomology
should be computed in the standard space of conformal fields that
we have been using.

(3) From the point of view of the Lagrangian \rpo, the symmetries
of $Q$ (volume preserving diffeomorphisms, say)
are a peculiar sort of big global symmetry group;
no gauge fields are in sight.  However, in string theory,
one believes that the unbroken symmetries at, say, the $SU(2)$ point
are the residue of a large underlying gauge group.  In particular,
perturbations of the $SU(2)$ point that are equivalent under
volume preserving symmetries of $Q$ (and other closed string symmetries
discussed presently) should be identified in the string theory;
but this is not automatic in \rpo.

\subsection{Dimensional Reduction}

The Lagrangian \rpo\ has been written in terms
of $H$ invariant quantities on $W$.  If we wish, we can carry out
explicitly the dimensional reduction to $Q$.  The abelian gauge
field $A$ on $W$ reduces  (after picking, non-invariantly, a section
of the bundle $W\to Q$) to $A=u\cdot dt + a$, where $a$ is the pullback
of an abelian gauge field from $Q$, $u$ is a function on $Q$, and $t$ is
a parameter along the fibers of $W\to Q$.\footnote{*}{$H$ invariance
of $a$ implies $(d i_S + i_S d)a=0$.  $a$ being a pullback means
$i_S a = 0$.}  $t$ is determined only up to
$t\to t-f$, where $f$ is a function on $Q$.  Under this
transformation, we have
$$\eqalign{ u & \to u \cr
            a & \to a - u \cdot df, \cr}\eqn\felvo$$
The Lagrangian when written in terms of these variables becomes
$$L=2\int_Q \sigma\cdot du\wedge da,      \eqn\riffox$$
which is invariant under \felvo.  We presently will understand better
the origin of this symmetry.

\section{Symmetries Of The Closed String}

To study the closed string symmetries, we must look at BRST invariant states
of ghost number one (more on this in \S4 ).
In keeping with our general analysis,
such a state $|\Psi\rangle$  represents
an  $H$ invariant vector field, $V=V^I\partial_I$
($I=1\dots 4$ runs over $x,y,x',y'$) on $W$.  Dually,
$|\Psi\rangle$ determines the three form
$\lambda=\epsilon_{IJKL}V^Idx^Jdx^Kdx^L
   $
on $W$.

Such a $|\Psi\rangle$ determines a symmetry of the closed string theory
if and only if $(b_0-\overline b_0)|\Psi\rangle =0$.  In keeping
with our general analysis, this translates into $d\lambda=0$ or,
dually,
$$\partial_IV^I=0 . \eqn\guffo$$
The latter condition means that $V$ generates a volume-preserving
diffeomorphism of $W$, that is a diffeomorphism that preserves the
volume form $dx\wedge dy\wedge dx'\wedge dy'$.

The symmetries found in [\witten] are determined by $\Psi$'s that
obey $b_0|\Psi\rangle=\overline b_0|\Psi\rangle=0$.  Such $\Psi$'s
correspond to generators of diffeomorphisms that preserve both
the left and right moving volume forms $\omega= dx\wedge dy$ and
$\omega'=dx'\wedge dy'$.  After imposing the $H$ invariance, they are
equivalent to volume preserving diffeomorphisms of $Q=W/H$.
The novelty now is that
by requiring  $\Psi$ to be annihilated only by $b_0-\overline b_0$,
we include diffeomorphisms that preserve $\omega\wedge\omega'$ but
do not preserve $\omega$ or $\omega'$ separately.

The extra symmetries can be described as follows.  The basic one is
the $H$ generator
$$ S=x{\partial\over\partial x}+y{\partial\over\partial y}
  -x'{\partial\over\partial x'}-y'{\partial\over\partial y'}\eqn\offfo$$
$S$ is clearly volume preserving, $\partial_IS^I=0$. It preserves
$\omega\wedge\omega'$ but not $\omega$ or $\omega'$.
More generally,
if $f$ is any $H$ invariant function, then $f\cdot S$ generates
a diffeomorphism that  is also volume preserving; indeed, $\partial_I(fS^I)
=S^I\partial_If=0$ by $H$ invariance of $f$.

How do these new symmetries act on the states?  $S$ annihilates all of them,
this being the condition of $H$ invariance, but this is not true of
$U=fS$.  Its action can be deduced from the
description of the states as $H$ invariant differential
forms on $W$.
The transformation of a differential form $\lambda$
under the symmetry generated
by a vector field $U$ is in general
$$\delta \lambda = {\cal L}_U \lambda =
i_U(d\lambda)+d(i_U\lambda) \eqn\tufgo$$
where $i_U$ is the operation of contraction of $U$ with one
index of $\lambda$.  The closed string states are closed differential
forms on $W$, so we can delete the $i_U(d\lambda)$ term in \tufgo.
Also, $H$ invariance of $\lambda$ means that \tufgo\ vanishes if $U=S$,
so $d(i_S\lambda)=0$.  In general, if $U=fS$, then $i_U\lambda
= f\cdot i_S\lambda$.  Hence \tufgo\
reduces for symmetries of this type to
$$\delta\lambda = df\wedge i_S\lambda.  \eqn\ufgo$$

The new symmetries act trivially on the ground ring.
If indeed $\lambda$ is a ghost number zero ring state, then
it corresponds to a four form, so $\delta\lambda$ is also a four form.
{}From $i_S(df)=0$ and $i_S^2=0$ it follows that $i_S\delta\lambda
=0$.  A four form annihilated by $i_S$ must vanish, so $\delta\lambda=0$.

In general, $\delta\lambda$ is given by a differential operator
of degree zero acting on $\lambda$; symmetries of this type are usually
called ``internal symmetries.''

The new symmetry action on the closed string moduli can be made very
explicit.  In fact, after making explicitly the dimensional reduction to
$Q$, the new symmetries correspond precisely to those described in \felvo.
This is no accident.  The new symmetries are generated by $H$ invariant
vector fields on $W$ which (as they act trivially on the ground ring)
project to zero on $Q$; thus they translate the section of $W\to Q$ that
was used in the dimensional reduction.

\section{Combining Left And Right Movers}

In this discussion, we have been building closed string states
by taking tensor products of left and right moving states and then
selecting states annihilated by $b_0-\overline b_0$.  We will now
discuss the justification for this more fully.  This subsection can
be omitted by readers willing to accept the procedure that we have followed
above on faith.

First of all, in considering the BRST cohomology, we can assume
that all states discussed below are
annihilated by $L_0$ and $\overline L_0$. The reason for this
is really that (i) $L_0$ and $\overline L_0$ generate a compact group over
which one can average; (ii) as $L_0$ and $\overline L_0$ are BRST commutators,
the cohomology class of a state is not changed in this averaging.

\REF\fgz{I. B. Frenkel, H. Garland and G. J. Zuckerman, Proc. Natl.
Acad. Sci. USA, {\bf 83}, (1986) 8442.}
\subsection{Review of Chiral BRST Cohomology}

To begin with, we review some facts about BRST cohomology in the chiral
case [\fgz].  There are two relevant kinds of BRST cohomology.
First, there is the ``absolute'' cohomology $H^n$; an element of $H^n$
is represented by a BRST invariant state $|\Psi\rangle$ of ghost number
$n$, modulo $|\Psi\rangle \to |\Psi\rangle +Q|\Lambda\rangle$ where
$|\Lambda\rangle$ has ghost number $n-1$.
Second, there is the relative cohomology
$H^n_R$.
An element of $H^n_R$ is represented
by a BRST invariant state $|\Psi\rangle$ of ghost number $n$, annihilated
by $b_0$, modulo $|\Psi\rangle\to |\Psi\rangle+Q|\Lambda\rangle$,
where $|\Lambda\rangle$ has ghost number $n-1$ and is annihilated by
$b_0$.

%What is relevant to closed string is really $H^n_S$, but what is most
%easily computed by combining left and right movers is $H^n$.
%Indeed, if $H^n_+$ and $H^n_-$ are the BRST cohomology of left and
%right movers, then
%$$H^n=\oplus_m H^m_+\otimes H^{n-m}_-. \eqn\gifox$$

There are standard maps between absolute and relative cohomology, that
fit together into an exact sequence.
First, there is a map
$$i:H^n_R\to H^n \eqn\hoggo$$
which consists of forgetting the $b_0$ condition.
Second, there is a map
$$b_0:H^n\to H^{n-1}_R \eqn\joggo$$
of multiplication by $b_0$ (which maps arbitrary states
to states annihilated by that operator, and preserves the $Q$ invariance
since we are working in the null eigenspace of $L_0$).
Finally there is a map
$$\{ Q , c_0 \}:H^n_R\to H^{n+2}_R, \eqn\woggo$$
that consists of multiplication by $\{Q,c_0\}$
(which is not a BRST commutator in the relative cohomology).
It is easy to see that these maps are all well defined.

These maps fit together into the ``exact sequence''
$$\dots H^n_R
\onarrow{i}
H^n \buildrel b_0 \over \longrightarrow
H^{n-1}_R \buildrel \{ Q,c_0 \} \over \longrightarrow
H^{n+1}_R \onarrow{i} H^{n+1}\dots .\eqn\xoggo$$
The statement that this is an exact sequence means (i) the composition
of two successive maps is zero; (ii) a state in the kernel of one map
is in the image of the one before.  The first statement is easy to see and
left to the reader;
the second will now be verified.

For instance, let us verify exactness of \xoggo\ at $H^n$.  An
element $w$ of $H^n$ is in the kernel of $b_0$ if any representative
BRST invariant state $|\Psi\rangle$
obeys $b_0|\Psi\rangle
=Q|\Lambda\rangle$ for some $|\Lambda\rangle $ which is annihilated
by $b_0$.  If so, pick for $w$ the representative
$|\Psi'\rangle =|\Psi\rangle
+{}Qc_0|\Lambda\rangle$; then $|\Psi'\rangle$
is annihilated by both $Q$ and $b_0$, so
represents a state $w'$ in $H^n_R$; and $w=i(w')$.

To verify exactness at
$H^{n-1}_R$, an element of the kernel
of $\{ Q,c_0\}$ can be represented by a state
$|\Psi\rangle$, of ghost number
$n-1$, annihilated by both $Q$ and $b_0$, and such
that $\{Q,c_0\}|\Psi\rangle=Q|\Theta\rangle$,
where $|\Theta\rangle$ is annihilated by $b_0$.
This being so, let $|\widetilde\Psi\rangle=c_0|\Psi\rangle
-|\Theta\rangle$; as $|\widetilde\Psi\rangle$ is annihilated by $Q$
and $|\Psi\rangle =b_0|\widetilde\Psi\rangle$, $ |\Psi\rangle$
is in the image of $b_0$, as we wished to show.

Finally, to show exactness at $H^{n+1}_R$,
suppose a state $|\Psi\rangle$ annhilated by $Q$ and $b_0$
is in the kernel of $i$.  This means that $|\Psi\rangle =Q|\Lambda\rangle$
(but $b_0|\Lambda\rangle$ may be nonzero, so the element
of $H^{n+1}_R$ represented by $|\Psi\rangle$ may be nonzero).
Any state $|\Lambda\rangle$ can be written
uniquely as $|\Lambda\rangle =c_0|\Lambda'\rangle
+|\Lambda''\rangle$ where $|\Lambda'\rangle$ and $|\Lambda''\rangle$
are annihilated by
$b_0$.  Since $|\Psi\rangle$ and $|\Psi\rangle
-Q|\Lambda''\rangle$ represent the same element of $H^{n+1}_R$, we can suppose
that $|\Lambda''\rangle=0$, so
$|\Psi\rangle=Qc_0|\Lambda'\rangle$,
with $|\Psi\rangle,\,\,|\Lambda'\rangle$ annihilated by $b_0$.
{}From $b_0|\Psi\rangle=0$, we learn that $Q|\Lambda'\rangle
=0$, a formula that implies $|\Psi\rangle=\{Q,c_0\}|\Lambda'
\rangle$, which now finally tells us that $|\Psi\rangle$ is in the image
of $\{ Q,c_0 \}$.

What we have said so far is true for arbitrary bosonic string vacua.
For the $D=2$ string vacuum we are studying, it is also true that
the map $\{ Q , c_0 \}$ is zero since it adds two units of ghost
number and relative cohomology exists
only for ghost numbers $0,1$ (plus states) and $2,3$ (minus states).
Since the map in question cannot mix plus and minus states it
follows that it must equal zero.

\subsection{Closed Strings}

In discussing closed strings, we consider models in which the total
Hilbert space is a tensor product of left and right moving Hilbert spaces
and the BRST operator $Q$ is a corresponding sum
$Q=Q_L+Q_R$, with $Q_L$ and $Q_R$ the BRST operators of left and right movers.

There are now several types of BRST cohomology
to consider.  The absolute closed string cohomology ${\cal H}^n$
consists of $Q$ invariant states of ghost number $n$, modulo $Q|\Lambda\rangle$
for arbitrary $|\Lambda\rangle$.  What we will call the
relative closed string cohomology
${\cal H}^n_R$ consists of $Q$ invariant states of ghost number $n$,
annihilated by $b_0$ and $\overline b_0$, modulo $Q|\Lambda\rangle$
where $|\Lambda\rangle$ is annihilated by $b_0$ and $\overline b_0$.
And what we will call the semirelative cohomology ${\cal H}^n_S$
consists of $Q$ invariant states of ghost number $n$ annihilated by
$b_0-\overline b_0$, modulo $Q|\Lambda\rangle$ where $|\Lambda\rangle
$ is annihilated by $b_0-\overline b_0$.
It is convenient to set $b_0^\pm=b_0\pm\overline b_0$, and
$c_0^\pm=c_0\pm \overline c_0$.

Let $H^n_+$ and $H^n_-$ be the absolute BRST cohomology of right
movers and left movers, respectively; and similarly let $H^n_{R,\pm}$
be the relative cohomology of right and left movers.  If these
are known, then ${\cal H}^n$ and ${\cal H}^n_R$ can be readily
extracted, since they are determined
by conditions that factorize between left and right movers.  In
fact
$$\eqalign{
{\cal H}^n & =\oplus_m H^m_+\otimes H^{n-m}_- \cr
{\cal H}^n_R& = \oplus_m H^m_{R,+}\otimes H^{n-m}_{R,-}. \cr}\eqn\uppo$$
But what we really need is ${\cal H}_S$, for which we must work a little
bit harder.

The various ${\cal  H}$'s are related by exact sequences analogous
to \xoggo.  First, one has
$$\dots {\cal H}^n_R \onarrow{i'}
{\cal H}_S^n \buildrel b_0^+ \over \longrightarrow
{\cal H}^{n-1}_R
\buildrel \{ \Q , c_0^+ \} \over \longrightarrow
{\cal H}^{n+1}_R \onarrow{i'} {\cal H}_S
^{n+1}\dots ,\eqn\uxoggo$$
where $i'$ is the map that forgets the extra $b_0$ condition, and
the next two maps are simply defined by multiplication with the
indicated operators.
Similarly, one has
$$\dots {\cal H}^n_S \onarrow{i''} {\cal H}^n
\buildrel b_0^- \over \longrightarrow
{\cal H}^{n-1}_S
\buildrel \{ \Q , c_0^- \} \over \longrightarrow
{\cal H}^{n+1}_S \onarrow{i''} {\cal H}
^{n+1}\dots ,\eqn\uuxoggo$$
where $i''$ is the map that forgets the $b_0^-$ condition and the
next two maps are defined by multiplication with the indicated
operators. The proof of exactness follows the discussion of
\xoggo\ line by line.

The formulas just written hold for arbitrary bosonic string vacua.
In the standard $D=2$ vacua, one has the further fact that in \uxoggo,
$\{Q,c_0^+\}$ vanishes.  The reason for this is that in the decomposition
${\cal H}_R^n=\oplus H_{R,+}^m\otimes H_{R,-}^{n-m}$, $\{Q,c_0^+\}$
would
increase $m$ or $n-m$ by 2, but $H_{R,\pm}^m$ is zero except for $m=0,1$
for plus states (and $m=1,2$ for minus states).
Hence \uxoggo\ reduces to exact sequences
$$0\to {\cal H}^n_R\onarrow{i'} {\cal H}_S^n\onarrow{b_0^+}
{\cal H}^{n-1}_R
\to 0. \eqn\topo$$
This means that as a vector space, ${\cal H}_S^n$ has a (non-canonical)
isomorphism
$${\cal H}_S^n\cong i'({\cal H}^n_R)\oplus {\cal H}^{n-1}_R. \eqn\loppo$$
This has a simple interpretation.  $i'({\cal H}^n_R)$ consists of closed
string states with representatives that are annihilated by both $b_0$
and $\overline b_0$.  These are the traditional closed string states,
the ones discussed for instance in [\witten].  \topo\ shows that the
other closed string states, which are annihilated by $b_0^-$
but not by $b_0^+$, are determined modulo $i'({\cal H}^n_R)$ by their image
under multiplication by $b_0^+$, this image being an arbitrary element of
${\cal H}^{n-1}_R$.  The recipe that we followed intuitively at the
beginning of this section to construct the closed string states can readily
be seen to agree with this: the reader can see in equations
\srpluscoh\ and \srminuscoh\ that at ghost number $n$ we have included
the states in the relative cohomology, plus extra states that are simply
obtained from the relative cohomology at ghost number $(n-1)$ upon
multiplication by $(a+\bar{a})$, which acts as an inverse for $b_0^+$.
We have therefore verified that this recipe is correct.

The recipe that we used is actually related in a more obvious way
to \uuxoggo.  If one knows that $\{Q,c_0^-\}=0$, then \uuxoggo\
implies that
$${\cal H}_S^n={\rm ker} \, b_0^-:{\cal H}^n\to {\cal H}_S^{n-1}.
\eqn\mormo$$
In other words, ${\cal H}_S^n$ consists of closed string states
annihilated by $b_0^-$.  That $\{Q,c_0^-\}=0$
depends on additional
facts (coming from explicit calculations of the BRST cohomology
[\lianzuck,\bouwknegt]).
\uxoggo\ is easier to work with, even though
it gives the result in a less transparent form, because the unknown
object ${\cal H}_S^*$ appears only once, and not twice as in \uuxoggo.

\chapter{The Closed String Symmetry Generators}

The purpose of the present section is the construction
of the generators of symmetries for the closed string
theory. As we will see, there are subtleties concerning
the proper combination of the left and right sectors of the theory.
Most of our attention will concentrate on the symmetries that
preserve total ghost number. One novel fact of these symmetries
is that they do not in general preserve both the left and the
right moving ghost numbers. There are also conserved charges that
change the total ghost number, and we will show how to
construct them.

For the purely holomorphic case the chiral symmetry algebra is
generated by the currents $J^\pm (z) = W^\pm_{s,n}(z)$,
which are dimension one, primary operators that give rise
to the charges $A_{s,n}$ defined by
$$A_{s,n}^\pm = {1\over 2\pi i} \CINT W^\pm_{s,n}(z) dz\eqn\chircurr$$
Since $W$ is built completely from the matter sector, one has
$\{ Q, W \} = \partial (cW)$, and as a consequence the above
charges map BRST invariant states to BRST invariant states
since
$$\{ Q, A \} = \CINT \{ Q, W \} dz = \oint \partial (cW) dz = 0.$$

Let us now turn to the case of the closed string. Conventionally
charges are built by integrating around a closed curve a
a holomorphic (1,0) operator or an antiholomorphic (0,1) operator.

In the present model, we can construct $(1,0)$ and $(0,1)$ operators,
but because of the requirement of matching the Liouville momenta,
they are not
holomorphic or antiholomorphic.
The natural way to
obtain a (1,0) operator is to multiply a purely holomorphic current
$W_{s,n}(z)$ by an antiholomorphic operator of dimension zero,
of zero ghost number and carrying the same Liouville momentum, in other
words an operator in the antichiral ground ring.  $(0,1)$ operators
are constructed similarly.  So we have [\witten]
$$\eqalign{
(1,0):\, \J_{s,n}(z,\bz ) &= W_{s,n}(z)\,\bO_{s-1,n}(\bz ),\cr
(0,1):\,\bJ_{s,n}(z,\bar{z}) &=  \O_{s-1,n}(z)\, \bW_{s,n}(\bz ).\cr}
\eqn\clcurr$$
These operators satisfy neither $\bpd \J =0$ nor $\pd \bJ = 0$.
In order to understand properly the nature of this complication let
us recall the general framework for conserved currents.

\section{Conserved Currents via Descent Equations}

In general a strictly conserved current corresponds
to a pair $(J_z,{J}_{\bz})$ such that the one form
$$\Omega^{(1)} =  J_z dz  - {J}_{\bz} d\bz ,\eqn\oneform$$
is closed
$$d \Omega^{(1)} = 0 \quad \leftrightarrow \quad
{\partial}_{\bar z}J_z + \partial_z {J}_{\bz} = 0.\eqn\consvcurr$$
The conserved charge $A$ is then given by
$$A({\cal C}) = \CINT\, \Omega^{(1)} =
\CINT\, dz J_z - \CINT \, d\bz {J}_{\bz},
\eqn\consvchar$$
where the two integrals are taken on the same contour and with the
same orientation. We take henceforth the orientation to be
such that
$$ \CINT \,{dz \over 2\pi i} {1\over z} =
-\CINT \,{d\bz \over 2\pi i} {1\over \bz} =1 .\eqn\convorient$$
Conservation is the statement that for two curves ${\cal C}$
and ${\cal C}'$ that are homologous, namely $\partial M = \C - \C'$
for some surface $M$, one has
$$0 = \int_M d \Omega^{(1)} = \int_{\partial M} \Omega^{(1)}
=\int_{\C} \Omega^{(1)}
-\int_{\C'} \Omega^{(1)} = A(\C ) - A(\C' ) .\eqn\strictcon$$

In BRST quantization, to have a conserved charge that is well defined
in the physical Hilbert space, we do not really need $A(\C)=A(\C')$.
It is enough for this to hold modulo BRST commutators,
$$A(\C')  = A(C)  + \{\Q , B \}.\eqn\modcharge$$
For then $A(\C')$ and $A(C)$ have the same action on physical states.
This possibility corresponds to the case when the
form $\Omega^{(1)}$ is closed only up to a BRST commutator.
Instead of \consvcurr\ we have
$$d\Omega^{(1)} = \{ \Q , \Omega^{(2)} \} ,\eqn\relaxcons$$
for some two-form $\Omega^{(2)}$. More explicitly, if
$\Omega^{(2)} = \Omega^{(2)}_{z\bz} \hbox{dz} \wedge d\bz
$, then $$
{\partial}_{\bar z}J_z + \partial_z{J}_{\bz} =
- \{ \Q , \Omega^{(2)}_{z\bz} \} .\eqn\ee$$
In this situation,
\strictcon\ is replaced with
$$A(\C ) - A(\C ') = \{ \Q , \, \int_M \Omega^{(2)} \}, \eqn\relaxch$$
as
in \modcharge.  This is good enough to give a conserved charge at the
physical level, but as we will see in \S6, Ward identities
can have unusual properties when $\Omega^{(2)}\not= 0$.

We must also formulate the condition for a conserved charge to commute
with $Q$.
This will happen if there is
a zero-form $\Omega^{(0)}$ such that
$$ d\Omega^{(0)} = \{ \Q , \Omega^{(1)} \} \eqn\entercoh$$
since in this case
$$\{ \Q , \CINT \,\Omega^{(1)} \}
=  \CINT\, \{ \Q , \Omega^{(1)} \}
= \CINT d\Omega^{(0)} = 0\eqn\zeroform$$
as desired.
It is then necessarily also true that
$\{Q,\Omega^{(0)}\}=0$.  Indeed,  \zeroform\ implies that $d\{Q,\Omega^{(0)
}\}=0$,
so $\{Q,\Omega^{(0)}\}$ is a $c$-number; this $c$-number must
vanish, or the identity would be a BRST commutator, and the whole
BRST machinery would break down.

Summarizing, the general framework for conserved
charges is given by the descent equations
$$\eqalign{ 0 & = \{Q,\Omega^{(0)}\}\cr
d\Omega^{(0)} &= \{ \Q , \Omega^{(1)}\} \cr
d\Omega^{(1)} &= \{ \Q , \Omega^{(2)} \}.\cr}\eqn\descent$$
These imply that $A= \CINT \,\Omega^{(1)}$ is a BRST invariant charge
conserved up to BRST trivial operators.

Let us close this section with a few comments.
To derive the symmetry charges
we have to choose BRST invariant zero forms $\Omega^{(0)}$. We
will use below the BRST cohomology classes. Since this model has
cohomology classes of several ghost numbers we can obtain symmetry
charges of various ghost numbers and different statistics. While
we will concentrate next on the case of the ghost number zero charges,
the other charges play a role on the larger symmetry structure
discussed in \S6 .

The solution of the descent equations is not unique.  If we have
one solution, we can generate another one via the replacements
$$\eqalign{
\Omega^{(0)} &\quad \rightarrow \quad \Omega^{(0)} +
\{ Q ,\alpha^{(0)} \} \cr
\Omega^{(1)} &\quad \rightarrow \quad \Omega^{(1)} + d\alpha^{(0)} +
\{ Q ,\alpha^{(1)} \} \cr
\Omega^{(2)} &\quad \rightarrow \quad \Omega^{(2)} + d\alpha^{(1)}
+\{ Q ,\alpha^{(2)} \} ,\cr}\eqn\gaugedescent$$
where $\alpha^{(i)}$ is an arbitrary form of degree $i$. It is
clear from these equations that the charge $\oint \Omega^{(1)}$ is
independent of the choice of BRST representative for $\Omega^{(0)}$.

The descent equations can be discussed effectively (and solved!) using
states. In terms of states the zero form $\Omega^{(0)}(z)$ correspond
to the state $\ket{\Omega^0} =\Omega^{(0)}\ket{{\bf 1}}$
(where $\ket{{\bf 1}}$ is the
SL(2,C) vacuum). Similarly, the one form
$\Omega^{(1)} = \Omega^1_z\, dz + \Omega^1_\bz\, d\bz$ defines the
states $\ket{\Omega^1_z} = \Omega^1_z \ket{{\bf 1}}$, and
$\ket{\Omega^1_\bz} = \Omega^1_\bz \ket{{\bf 1}}$. Finally
$\Omega^{(2)} = \Omega^2 dz\wedge d\bz $ gives rise to the
state $\ket{\Omega^2} = \Omega^2 \ket{{\bf 1}}$. If we have
a suitable energy momentum tensor $(T(z), \overline T (\bz))$ such
that  $\ket{\partial \O} = L_{-1} \ket{\O}$ and
$\ket{\bar\partial \O} = {\overline L}_{-1} \ket{\O}$ then we can
write the descent equations as:
$$\eqalign{
0\,\,\, &= Q \ket{\Omega^0} , \cr
L_{-1} \ket{\Omega^0} &= Q \ket{\Omega^1_z}\cr
{\overline L}_{-1} \ket{\Omega^0} &= Q \ket{\Omega^1_\bz}\cr
L_{-1} \ket{\Omega^1_\bz} - {\overline L}_{-1} \ket{\Omega^1_z}
&= Q \ket{\Omega^2} .\cr}\eqn\statesdescent$$
Since the Virasoro operators can be written as
$L_{-1} = \{ Q , b_{-1} \}$ and
${\overline L}_{-1} = \{ Q , \bar{b}_{-1} \}$
it follows that the above equations are readily solved by
$$ \ket{\Omega^1_z} = b_{-1} \ket{\Omega^0}, \quad
 \ket{\Omega^1_\bz} = \bar{b}_{-1} \ket{\Omega^0}, \quad
 \ket{\Omega^2} = b_{-1} \bar{b}_{-1} \ket{\Omega^0}.\eqn\statessol$$
Again, the descent equations for the states
do not give a unique solution. Other solutions can be obtained
using the analog of \gaugedescent\ for states.

\section{Currents of Ghost Number Zero}

\REF\siegel{W. Siegel, Introduction to String Field Theory (World
Scientific, Singapore, 1988) .}
\REF\npcsft{M. Saadi and B. Zwiebach, Ann. Phys. {\bf 192} (1989) 213;
\hfill\break
T. Kugo, H. Kunitomo and K. Suehiro, Phys. Lett. {\bf B226} (1989) 48;
\hfill\break
T. Kugo and M. Suehiro, Nucl. Phys. {\bf B337} (1990) 434}
\REF\npcsftl{B. Zwiebach, Mod. Phys. Lett. {\bf A5} (1990) 2753;
\hfill\break
B. Zwiebach, Comm. Math. Phys. {\bf 136} (1991) 83; \hfill\break
H. Sonoda and B. Zwiebach, Nucl. Phys. {\bf B331} (1990) 592}

We will now derive the currents and associated charges of
ghost number zero by using the descent equations. In order
to motivate the choice we will make for the zero-forms, it
is useful to understand how symmetry charges arise
in string field theory [\siegel ]. In the BRST closed string field
theory [\npcsft , \npcsftl ] the gauge symmetries are written as
$$\delta b_0^- \ket{\Psi} = \Q \, b_0^-\ket{\Lambda}
+ g \ket{\Psi \star \Lambda} + \cdots \eqn\sftgauge$$
where the dots represent higher order contributions due
to the nonpolynomiality of the theory.
Here $b_0^- \ket{\Psi}$, the string field,
is an element of
the Hilbert space of ghost number (+2) (in the conventions
that the ghost number of the SL(2,C) vacuum is zero)
annihilated by $b_0^-$, and $b_0^-\ket{\Lambda}$,
the gauge parameter, has ghost number (+1) and is also
annihilated by $b_0^-$. Unbroken symmetries correspond
to transformations for which $\Q b_0^- \ket{\Lambda} = 0$.
In this case, to first order, the symmetry acts on the
string field linearly, in a manner determined by the
gauge parameter and the string product.
A BRST trivial state $b_0^- \ket{\Lambda}$, however,
will generate a symmetry which vanishes on-shell.
Thus the nontrivial unbroken symmetries correspond to the
closed string BRST semi-relative cohomology classes
at ghost number $(+1)$. This is a general statement in
string field theory. The existence of these cohomology classes
must explain the origin of the symmetry transformations we are
looking for. The relation is actually very simple: the BRST
classes simply give us the zero forms $\Omega^{(0)}$ that determine
the charges!

We can now understand the current
whose $(1,0)$ piece is ${\cal J}_{s,n}=W_{s,n}\overline{\cal O}_{s-1,n'}$.
(The analysis of the operator whose $(0,1)$ piece is ${\cal O}\overline
{\cal J}$ is of course precisely analogous.)
In this subsection, we will suppress the subscripts $s,n,n'$.
The current we are seeking will arise from the zero form
$$\Omega^{(0)}= cW\overline{\cal O}. \eqn\mimoc$$
To find the higher components of this operator, we need the
chiral descent equations for $cW$ and $\overline {\cal O}$.
For $cW$, we have
$$\eqalign{ 0 & = \{Q_L,cW\}=\{Q_R,cW\} \cr
  {}& \{Q_L,W\} = \partial(cW)\cr
      0 & =\{Q_R,W\}= \bar\partial (cW).  \cr}\eqn\piip$$
Of course, $Q_L$ and $Q_R$ are the holomorphic and antiholomorphic parts
of the BRST operator; thus $Q=Q_L+Q_R$.
Similarly,
$$\eqalign{ 0 & = \{Q_L,\overline{\cal O}\}=
\{Q_R,\overline{\cal O}\} \cr
   0 &=\partial\overline{\cal O} \cr
  {}& \{Q_R,\bar X\}=\overline\partial \,\,\bar{\cal O}, \cr} \eqn\iip$$
for some
spin $(0,1)$ operator $\bar X$.  In terms of states it is simply
given by $\ket{\overline{X}}$
$= \bar{b}_{-1} \ket{\overline{\O}}$. This is actually a highest
weight state, and the corresponding operator $\bar X$ is primary.
This is the case (see the discussion at the beginning of \S2.2 )
since $\ket{\overline{\O}}$ is annihilated by $\bar{b}_n$ for all
$n\geq 0$ (Appendix A).

Using \piip\ and \iip, one readily constructs the higher components
of $\Omega^{(0)}$:
$$\eqalign{
\Omega^{(1)} & = W\overline{\cal O} dz -cW\overline Xd\bar z  \cr
\Omega^{(2)} & = -W\overline X \,\,dz\wedge d\bar z. \cr}
\eqn\gumob$$
The current that is conserved up to $\{Q,\dots\}$ thus has
$(1,0)$ component $W\overline{\cal O}$, as expected.  In addition,
it has a $(0,1)$ piece $cW\overline X$, which may be less expected.
Restoring the indices, the conserved charges are thus
$$\eqalign{\A_{s,n,n'} &= {1\over 2\pi i}
\CINT\, dz\, W_{s,n} \bO_{s-1,n'}
- {1\over 2\pi i} \CINT\, d\bar{z} \, cW_{s,n} \bX_{s-1,n'}.\cr}
\eqn\clchar$$
In \S5.3, we do an explicit computation showing that the second piece
is needed to obtain sensible (BRST invariant) results when these
charges act on discrete moduli.

As we will see in \S6, the two-form $\Omega^{(2)}$ derived from
the zero-form $\Omega^{(0)}$ will be responsible for the
nonlinear action of the symmetry transformations on states.
In terms of states the two form given in \gumob\ is indeed given
by
$$ \ket{\Omega^2} \, = \, b_{-1} \bar{b}_{-1} \ket{\Omega^0}.
\eqn\descentst$$
as predicted by our analysis leading to equation (\statessol ).
Equation \descentst\ gives a simple test for nonlinear
symmetries; if the two form vanishes the symmetry will
act linearly on states.

\section{Charges Preserve the $b_0 - \bar{b}_0 = 0$ Condition}

As we have discussed in \S3, physical states $\ket{\psi}$,
in addition to being annihilated by the BRST operator, must
satisfy the condition
$$ (b_0 - \bar{b}_0) \ket{\psi} = 0,
\eqn\physcond$$
where
$$b_0-\bar{b}_0 = b_0^- = \CINT {dz\over 2\pi i} zb(z)
+\CINT {d \bz\over 2\pi i} \bz \bar{b}(\bz ),\eqn\defbminus$$
and the contour surrounds the origin.
We now want to verify that if $\ket{\psi}$ is acted
on by a charge ${\cal A}_{s,n,n'}$, it still satisfies \physcond.
(One should also verify this for the new symmetries that
we have constructed, but we have not undertaken this.)
We must then check that
$[b_0-\bar{b}_0 ,{\cal A}_{s,n,n'} ] = 0$. Using \clchar\ with the
integration label $z$ changed into $w$ and with the contour chosen
to be a circle around the origin ($w=0$) we find
$$\eqalign{
[b_0-\bar{b}_0 ,{\cal A}_{s,n,n'} ] &=
\CINT {dw\over 2\pi i}\, W_{s,n}(w) \,\oint {d\bz \over 2\pi i}
\,\bz \bar{b}(\bz )\, \bO_{s-1,n'}(\bar{w}) \cr
{}&- \CINT {d\bar{w} \over 2 \pi i}\, w W_{s,n} (w)
\,\overline{X}_{s-1,n'}(\bar{w})\cr
{}&+ \CINT {d\bar{w} \over 2 \pi i} \,c(w) W_{s,n}(w)
\oint {d\bz \over 2 \pi i}\, \bz \bar{b} (\bz )
\,\overline{X}_{s-1,n'}(\bar{w}),\cr}\eqn\calculcom$$
where $\CINT$ refers to the circle around the origin in the
$w$ plane, and
$\oint$ is an integral around the $w$ points.  (In writing the first
term in \calculcom, we have used the fact that the
$W\overline{\cal O}$ term in the charge can only be acted upon by
$\bar{b}_0$ since $W$ is constructed without ghosts.)
Vanishing of \calculcom\ can be established using
the explicit expression for $X_{k,k}(z)$ given in
Appendix A, which  shows that it is built with matter
and antighost fields only; it contains no ghost fields.
Using the SU(2) lowering operators, the same must
hold for $X_{s,n}$ and of course also for $\bar X_{s,n}$.
Thus the last term in the above
equation vanishes as there is no short distance singularity
in the operator product expansion of $\bar{b}$ and
$\overline{X}$. It also follows from Eqn. (A.17) that
$$b(z) \,\O_{s,n}(w) = {1\over z-w} X_{s,n}(w) + \cdots \eqn\opebo$$
Using this result for the first term in the right hand side
of \calculcom\ we then find
$$[b_0-\bar{b}_0 ,{\cal A}_{s,n,n'} ] = -\CINT
\left( \bar{w}\, {dw\over 2 \pi i} + w\, {d\bar{w} \over 2 \pi i}
\right) W_{s,n} \overline{X}_{s-1,n'} = 0\eqn\finalcom$$
where in the last step we used the fact that the contour was a circle,
on which $|w|^2={\rm constant}$ so $\bar w dw+wd\bar w=0$.
This establishes the desired result.

There is one more point to be discussed. The descent equations show
that upon contour deformation a conserved charge acquires
an additional term of the form $\{ Q, \Omega^{(2)} \}$ integrated
over the part of the surface bounded by the contours. Thus acting
on a physical state $\ket{\psi}$ the additional term we get is
$$ Q \, \int \Omega^{(2)} \, \ket{\psi} = - Q \int dz\wedge d\bar z
\,\, W\overline X \, \ket{\psi} ,\eqn\checktriv$$
where we made use of \gumob.
While this term is clearly trivial in the absolute cohomology
we must show it is trivial in the semi-relative cohomology. It is
so.  The state $\ket{\psi}$ can be written as $b_0^- \ket{\chi}$,
and $b_0^-$ commutes with $W\overline X$ (recall both $W$ and
$\overline X$ do not contain the ghost field $c$); therefore the
above state is of the form $Q\, b_0^- \cdots \ket{\chi}$, which
is trivial in the semi-relative cohomology. Thus the charges
are well-defined on the physical states. The attentive reader
will have noticed that we made implicit use of this result in
choosing a convenient contour for the charge ${\cal A}$ in this
section.

\section{Discussion Of New Moduli and Symmetry Transformations}

In \S3, we found new closed string states, annihilated by $Q$ and by
$b_0-\overline b_0$, at ghost number 1 and 2.  Explicitly, they are
$$\eqalign{
G=1 &: \,\, (a+ \bar{a}) \O_{u,n} \bO_{u,n'} \cr
G=2 &: \,\, (a+ \bar{a}) Y^+_{s,n} \bO_{s-1,n'}, \quad
(a+ \bar{a}) \O_{s-1,n} \overline{Y}^+_{s,n'}, \cr}
\eqn\newstuff$$
where $G$ is the ghost number. (We recall that $Y$ is the same as $cW$.)
We will first discuss briefly
the problem of explicitly describing the operators corresponding to
the new moduli, and then we will give the symmetry charges
associated with the states at $G=1$.

We recall that the moduli come from spin zero operators of ghost number two.
For the case of the ghost number two
states annihilated by $b_0$ and $\bar{b}_0$, the transition from
the states to the moduli (dimension (1,1) operators
that can be added to the action of the conformal field theory)
was simple. The states were of the type
$Y^+_{s,n} \bY^+_{s,n'}$ which are just
$cW^+_{s,n}\bar{c}\bW^+_{s,n'}$; the corresponding moduli are
$W^+_{s,n}\bW^+_{s,n'}$ which are simply obtained by deleting
the ghost fields.

As for the new states, the first nontrivial examples are given by
$$\eqalign{
(a+\bar{a}) (c\partial X \cdot 1 ) =
c\partial X \,(\bar{\partial} \bar{c} +
{1\over \sqrt{2}} \bar{c} \bar{\partial} \phi )
+  c \partial c \, \partial X, \cr }\eqn\firsttwoex$$
and similarly with left and right movers reversed.
It is clear that the simple rule to obtain the corresponding
moduli does not apply. Nevertheless, the $c\bar c \partial X\bar\partial \phi$
term in \firsttwoex\ must
correspond to a piece in the modulus of the form
$$\eqalign{
{}&\partial X \, \bar{\partial} \phi + \cdots .\cr
}\eqn\findops$$

This is not satisfactory as it stands, since the
field $\bar\partial \phi$ is not primary, because of
the background charge of the Liouville field, nor it is Weyl invariant.
The $c\bar\partial\bar c \partial X$ term
in \firsttwoex\ correspond roughly to an additional contribution
$\partial X\overline \omega$, where $\overline \omega$ is the $(0,1)$ part of
the spin connection.
With the right coefficient, this term restores Weyl invariance at the cost
of
local Lorentz invariance.  To save the situation, one must
take due account of the $c\partial c\partial X$ term.  This term enters
in verifying the
$b_0-\overline b_0$
condition and so is bound to enter in constructing the appropriate marginal
operator.
(New contributions to the local Lorentz and Weyl transformation laws
of $X$ may also be part of the story.)
We do not understand how to incorporate it in constructing
the operator, and so we leave this matter for later consideration.

\bigskip
Let us now turn to the new symmetries. In order to find the
expressions for the currents corresponding to the new
symmetries, we apply the descent equations.
The beginning point is the zero form corresponding
to the new BRST class of operators at $G=1$ (eqn. \newstuff )
$$\Omega^{(0)} = (a+ \bar{a}) \O_{u,n} \bO_{u,n'} =
a\O_{u,n} \bO_{u,n'} + \O_{u,n} \, \bar{a}\bO_{u,n'}.
\eqn\begcalc$$
We will use the notation
$$\partial \O = \{ Q, X_\O \} ,\quad
\partial (a\O ) = \{ Q, X_{a\O} \} ,\eqn\newnota$$
where $X_\O$ was simply denoted as $X$ before, and $X_{a\O}$
is an operator that actually fails to be primary, since
$a\O$ is not annihilated by $b_0$ (recall that as
for $X$, we have that $\ket{X_{a\O}} = b_{-1} \ket{a\O}$).
The formulas we are constructing here for the new symmetry charges
are therefore only adequate on the cylinder (flat metric).  Maintaining
current conservation in general
will involve adding to the currents additional terms involving coupling
to the world-sheet curvature; we do not know an efficient way to generate
these terms.

A short calculation with the descent equations gives us
$$\Omega^{(1)} = (X_{a\O} \bO + X_\O \bar{a}\bO ) dz
+ (\O \bX_{\bar{a}\bO} - a\O \bX_{\bO}) d\bz\eqn\oneformnew$$
$$\Omega^{(2)} = (X_\O \bX_{\bar{a}\bO} - X_{a\O}\bX_{\bO})
\, dz \wedge d\bz ,\eqn\twoformnew$$

The simplest charge is that corresponding to
$\Omega^{(0)} = (a + \bar{a}) \O_{0,0} \bO_{0,0}$
$= (a+ \bar{a})\, {\bf 1}\cdot {\bf 1}$. We then have
$X_{\bf 1} = 0$, and $X_{a\cdot {\bf 1}} = \partial \phi$.
It therefore follows that
$$\Omega^{(1)} = \partial \phi \, dz
+ \bar\partial  \phi \, d\bz,\eqn\simplestex$$
and recalling our sign convention \convorient\ we conclude that
the corresponding charge simply measures the difference between
left and right components of Liouville momentum. This is
the symmetry generator we denoted as $S$ in \offfo , and which
actually annihilates all states. The two-form corresponding to
this symmetry vanishes.  The other ``new'' charges, as discussed in
\S3.3, correspond  to vector fields $fS$, and have a nonvanishing
action.

\chapter{Symmetry Transformations of Tachyons and Discrete States}

In this section we show explicitly the linearized action
of the symmetry transformations on the states of the theory.
In the case of the uncompactified theory we will pay particular
attention to the transformation of the tachyon. The symmetry
generators which map the tachyon states to themselves generate
the subalgebra of the Virasoro algebra corresponding to the
$L_n$'s with $n\geq 0$. The transformation law of the tachyon
under these generators will turn out to show
that it is an object of dimension one.
In showing this, we will need to make a suitable
rescaling of the tachyon field  $\T_p$ with a momentum dependent
factor. It is very interesting that this factor coincides with
the external leg factor appearing in $S$-matrix calculations
of the $c=1$ theory. Thus, if the tachyon is defined to have
standard Virasoro transformations under the discrete symmetries,
the external leg factors go away from $S$-matrix elements.
We also give an alternative derivation of the transformation
law of the tachyon using the picture of tachyon states as
perturbations of the Fermi surface in the matrix model.
Finally, we give examples where we compute discrete symmetry
transformations of the discrete states. These examples illustrate
clearly what would have gone wrong with the naive charges, and the necessity
of including new states in the semirelative cohomology.

\section{Transformation of the Tachyon}

The calculation of the action of the discrete symmetries on
the holomorphic tachyon is useful in preparation for the
closed string case, so we consider this case first.

\noindent
$\underline{\hbox{The holomorphic case}}$.
Consider the holomorphic part of a tachyon on the right
branch of the spectrum $( p \geq 0)$
$$T_p (z) = e^{ipX(z)} \, e^{(\sqrt{2}- p)\phi (z)}\eqn\tachdef$$
It is clear from the momentum dependence of this field that
the momentum dependence of a symmetry generator that
maps a tachyon into a tachyon must be of the form
$\exp (iaX^+ )$ where we have defined $X^\pm = (X\pm i\phi )/\sqrt{2}$.
Since the $W^+_{s,n}$ operators are of the form
$$W^+_{s,n} \sim (\hbox{Polynomial in}\, \partial X)\,
e^{i(n+s-1)X^+}\, e^{i(n-s+1)X^-},\eqn\momdep$$
the relevant operators are the
ones generated by the currents $W_{s,s-1}$. We find an
explicit expression for these generators by acting with
the SU(2) lowering operator $\oint dz\,\,e^{-iX\sqrt 2}$
on $W_{s,s}$.  Thus
$$W_{s,s-1} = {1\over 2\pi i} \oint dz
e^{-iX\sqrt{2}} \,
e^{i\sqrt{2}sX(w)} \, e^{-\sqrt{2}(s-1)\phi (w)},$$
{}from which it follows that
$$W_{s,s-1} (w) = S_{2s-1} (-i \sqrt{2} X^{(j)}/j!)\,
e^{i\sqrt{2}(s-1)X^+(w)}\eqn\getwss$$
where $S_k$ denotes an elementary Schur polynomial (for definition
and notation see the Appendix)
and $X^{(j)} \equiv \partial^{(j)} X$. We are now ready to compute
the action of this generator on the tachyon
$$\{ A_{s,s-1} \, , T_p(w) \} = \res\oint W_{s,s-1}(z) T_p (w)$$
The contraction of the exponentials in $W$ and $T_p$ give a zero
$(z-w)^{2s-2}$ and this implies that the complete Schur polynomial
in $W$ must be contracted with the exponential in $T_p$ in order
to give a contribution. A small calculation gives
$$\{ A_{s,s-1} \, , T_p(w) \} = S_{2s-1} (\vec{c} \sqrt{2}p)
\, T_{p+\sqrt{2}(s-1)} (w) ,\eqn\troptach$$
where $c_k = (-1)^k/ k$.
It is possible, for our particular coefficients $c_k$,
to simplify the expressions of the Schur polynomials
down to factorials:
$$
S_k (\vec{c} a ) =
{(-1)^k \over k!} {\Gamma (a+k) \over \Gamma (a) },
\eqn\symsxxc$$
as derived in the Appendix. We can now write our result \troptach\ as
$$\{ A_{s,s-1} \, , T_p(w) \} = {(-1)^{2s-1}\over (2s-1)!}
{\Gamma (\sqrt{2}p + 2s-1) \over \Gamma (\sqrt{2}p)}
\, T_{p+\sqrt{2}(s-1)} (w) .\eqn\gammatach$$
The normalization of the operators $A_{s,s-1}$ has not been fixed yet.
We fix it by introducing
the operators $Q_{2s}$ via the relation
$$A_{s+1,s} = {(-1)^{2s} \over (2s+1)!} Q_{2s}.\eqn\fixnorm$$
Using \gammatach\ and \fixnorm\ the transformation for the tachyon
now takes the form
$$\{ Q_{2s} , T_p (w) \} = - {\Gamma (\sqrt{2}p + 2s + 1)
\over \Gamma (\sqrt{2} p )}\, T_{p+ {2s\over \sqrt{2}}}.\eqn\nicertr$$
The operators $Q_{2s}$ for $s= 0,{1\over 2}, 1, {3\over 2}, \cdots $,
generate half of the Virasoro algebra. Indeed  one readily verifies
that acting on the tachyon the $Q_{2s}$ operators satisfy the
commutation relations
$$\{ Q_{2s} \, , Q_{2s'} \} = (2s-2s') Q_{2s+2s'}.\eqn\virasoro$$

The transformation indicated in \nicertr\ is not yet of the
standard form. To this end we must redefine the tachyon by
a momentum dependent factor. Letting
$$T_p = {\hat T_p \over \Gamma (\sqrt{2} p + 1 )},\eqn\newtach$$
the transformation law for the tachyon becomes
$$\{ Q_{2s} , \hat{T}_p (w) \} = - \sqrt{2} p \,
\hat{T}_{p+ {2s\over \sqrt{2}}} (w).\eqn\supnicetr$$
This is the simplest form for the transformation law.
Consider now the standard transformation law for a field of
dimension $d$ under Virasoro:
$$\{ Q_{2s} , \phi _n \} = [2s(d-1) - n] \phi_{n+2s}.\eqn\standvir$$
The absence of an $s$ dependent term in the right hand side of
\supnicetr\ indicates that the tachyon transforms as a field of dimension
one under the symmetry transformations. The transformation
mixes tachyons whose momenta differ by an integer times $1/\sqrt{2}$.

\noindent
$\underline{\hbox{The Closed String Case}}$.
The field we want to study now is the closed string tachyon. In fact,
we must consider the BRST invariant tachyon field. While in the
holomorphic case there was no necessity to deal with the BRST
invariant field because the symmetry transformations do not involve
the ghosts, the closed string charges involve the ghost fields in
a nontrivial way. We must therefore consider the transformation of
$$\eqalign{
\T_p (z,\bz ) &= c(z) T_p(z) \, \bar{c} (\bz ) \bT_p (\bz ) ,\cr
{}&= c(z) e^{ipX(z)} \, e^{(\sqrt{2}- p)\phi (z)}
\,\,\bar{c}(\bz ) e^{ip\bX (\bz )} \, e^{(\sqrt{2}- p)\bar{\phi}
(\bz )}.\cr}\eqn\ctachdef$$
The operators that map tachyons into tachyons in this case are
$$\A_{s,s-1} = {1\over 2\pi i} \oint dz\, W_{s,s-1} \bO_{s-1,s-1}
- {1\over 2\pi i} \oint d\bar{z} \, cW_{s,s-1} \bX_{s-1,s-1}\eqn\gencl$$

We now wish to calculate $\{ \A_{s,s-1} , \T_p \}$. Let us begin
with the first term in the right hand side of \gencl . Recall from
the holomorphic calculation that the operator product expansion
of $W(z)$ with the holomorphic part of the tachyon $T_p(w)$ gave terms
of the form $(z-w)^{-1} , 1, (z-w), \cdots $. This implies that only
the regular terms in the antiholomorphic product of operators can
contribute. We therefore have to compute the regular terms in
$$\bO_{s-1,s-1} (\bz ) \,\,
\bar{c}\, \bT_p (\bar{w} ).\eqn\desired$$
It is actually sufficient to determine the regular terms up to
BRST commutators of the form $\{ \bQ , \star \}$. This is the
case because the holomorphic part of the calculation gave a
$Q$-invariant tachyon field, and therefore the antiholomorphic
BRST commutators can be recast as closed string BRST commutators.
We will use the properties of the ring to perform this calculation.
Note that since the momentum factors are the same as in the
holomorphic calculation we get a zero of the form
$(\bz - \bar{w})^{2s-2}$ that must be cancelled in order to get
a contribution. Since $\bO_{s-1,s-1} \sim {\bar{x}}^{2s-2}$ , where
$$\bar{x}=\bO_{{1\over 2},{1\over 2}}
= [\bar{c} \bar{b} + i \bar{\partial}\, \bX^- ]
e^{i\bX^+}$$
we first compute
$$\bar{x} (\bz ) \,  \,
\bar{c} \,\bT_p (\bar{w} ) \sim  (-\sqrt{2} p)\,\bar{c}\,
\bT_{p + {1\over \sqrt{2}}}(w)
+ \O (\bz -\bar{w}).\eqn\basicr$$
It is now straightforward to complete the calculation
$$\eqalign{
\bar{x}^n (\bz ) \,  \,
\bar{c} \,\bT_p (\bar{w} ) &\sim
(\bar{x}\cdot \bar{x} \cdots \bar{x}) (\bar{c}\,\bT_p)\cr
{}&=\bar{x}\cdot (\bar{x} \cdots
(\bar{x} \cdot (\bar{c}\,\bT_p ))\cdots )\cr
{}& = (-)^n (\sqrt{2}p)(\sqrt{2}p+1)\cdots (\sqrt{2}p+ n-1)
\,\bar{c} \,\bT_{p+n/\sqrt{2}}(\bar{w} ) \cr
{}&= (-)^n {\Gamma (\sqrt{2}p + n)\over \Gamma (\sqrt{2}p )}
\,\bar{c} \,\bT_{p+n/\sqrt{2}}(\bar{w}) + {\cal O}(\bz-\bar{w}).
\cr}\eqn\ringcomp$$
where in the second step we used associativity of the operator
product expansion, and then we made repeated use of \basicr\ .
We have therefore obtained the following result
$$\bO_{s-1,s-1} (\bz ) \,\,
\bar{c}\, \bT_p (\bar{w} )
\sim (-1)^{2s-2} {\Gamma (\sqrt{2}p + 2s-2)
\over \Gamma(\sqrt{2}p ) } \bar{c}\,\bT_{p+ (2s-2)/\sqrt{2}}
+ \O (\bar{z} - \bar{w}).\eqn\hereitis$$
The reader may enjoy verifying that the same result is obtained
by direct calculation using the representative for $\O_{k,k}$
given in (A.9) and equations (A.8) and (A.5).

We must now consider the second term in the charge $\A_{s,s-1}$,
namely,
$$- {1\over 2\pi i} \oint d\bar{z} \, cW_{s,s-1} \bX_{s-1,s-1}.$$
The holomorphic calculation involved here is essentially a copy
of the previous one since we just have an extra $c(z)$ factor.
This factor implies that this time we can only get terms regular
or vanishing in $(z-w)$. Therefore, in order to get a contribution,
the antiholomorphic contraction
$$\bX_{s-1,s-1} (\bz ) \,\bar{c}\,\bT_p (\bar{w}) $$
must give at least a first order
pole in $(\bz - \bar{w} )$.  The contraction of the exponentials
again gives us the zero $(\bz -\bar{w})^{2s-2}$. From the expression
for $X_{k,k}$ in the Appendix we see that the generic term in
$\bX_{s-1,s-1}$ is of the form $S_{2s-2-q}(\bar{\partial}\,\bX^-)$
$\partial^{q-1}b$ (times an exponential). The contractions with
the ghost factor and the exponential in the tachyon will therefore
give the pole $(\bz-\bar{w})^{2s-2}$. This is not singular enough
to yield a residue, and we therefore conclude that the second
term in the charge does not contribute in the present case.

Thus summarizing, we have from \gammatach\ and \hereitis\ that
the symmetry transformation on the tachyon reads:
$$\{ \A_{s,s-1} ,\T_p \} = {(-1)\over (2s-1)!}
{\Gamma (\sqrt{2}p + 2s-1) \over \Gamma (\sqrt{2}p)} \,
{\Gamma (\sqrt{2}p + 2s-2) \over \Gamma(\sqrt{2}p ) }
\T_{p+ (2s-2)/\sqrt{2}}.\eqn\thissofar$$
As in the holomorphic case, we introduce charges
$Q_{2s}$ which are suitably normalized, via the
relation
$$\A_{s+1,s}  = {(-1)^{2s}\over (2s+1)!}Q_{2s}.\eqn\normprop$$
(The sign factor has been chosen to agree with standard conventions.)
We then obtain the transformation law
$$\{ Q_{2s} , \T_p \} = - (-1)^{2s}
{\Gamma(\sqrt{2}p + 2s +1)\over \Gamma (\sqrt{2}p )}\,
{\Gamma(\sqrt{2}p + 2s)\over \Gamma (\sqrt{2}p )}\,
\T_{p+2s/\sqrt{2}}.\eqn\closedstach$$
It is straightforward to verify that, acting on the closed string
tachyon, the above operators
$Q_{2s}$ with $s=0,{1\over 2}, 1,\cdots $, form a subalgebra of
Virasoro.

It is possible to redefine the tachyon field $\T_p$ with
suitable momentum dependent functions, so as to bring the
above representation of Virasoro into a standard form.
We introduce a new tachyon field $\hat{\T}_p$ via:
$$\T_p = {\Gamma (1-\sqrt{2} p)  \over \Gamma (\sqrt{2}p )}
\,\,
{  {\hat{\T}}_p  \over \sqrt{2}p } ,\eqn\comparemm$$
where the first factor in the right hand side is precisely
the external leg factor of $S$ matrix elements in the
$c=1$ model and contains all
of the expected poles at discrete momenta (\cf\ the second
paper in [\grossnewman]). We then obtain
$$\{ Q_{2s} , {\hat{\T}}_p \} = -\sqrt{2} p \,
{\hat{\T}}_{p+2s/\sqrt{2}} .\eqn\okntransf$$
This is the final form for the transformation and proves that the field
whose $S$ matrix elements do not show discrete leg poles
transforms as a dimension one field under Virasoro.  The same would
be true for any field differing from this one by a momentum dependent
factor $h(p)$ satisfying the periodicity condition
$h(p+1/\sqrt{2}) = f_0 h(p)$, with $f_0$ a constant. This follows from
\okntransf\ and the fact that the Virasoro commutation relations
are invariant under the replacement
$Q_{2s} \rightarrow (f_0)^{2s} Q_{2s}$.

\section{Comparison To The Matrix Model}

Now we want to compare this result to the prediction of the $c=1$ matrix
model.

In the matrix model, the tachyon is described as a curve in the
phase space of the matrix eigenvalue, representing the fermi surface.
 We will call this phase
space the $a_1-a_2$ plane.
At zero cosmological constant, the fermi surface is the curve $a_1a_2=0$.
This has two branches, and we will look at the branch $a_1=0$.
A small deformation perturbs this equation to an
equation
$$ a_1-f(a_2)=0               \eqn\yillo$$
where $f(a_2)$ is the tachyon field.
Symmetries of the matrix model are generated by appropriate time dependent
canonical transformations.  For our purposes, we can ignore the time dependence
and consider charges acting at $t=0$.  The symmetries are then
transformations of the $a_1-a_2$ plane that preserve the area form
$$\omega=da_1\wedge da_2 .\eqn\illo$$
Such transformations are generated by Hamiltonian functions
$h(a_1,a_2)$ in the standard fashion.  As explained in [\witten,
\S3.1], the symmetries of the matrix model are generated not by arbitrary
$h$'s, but just by those $h$'s that vanish on the fermi surface $a_1a_2=0$
defining the vacuum state.  Since we are working near $a_1=0$,
we simply restrict $h$ to be divisible by $a_1$.

First, we will work to leading order near $a_1=0$.  This means that
we consider $h$ to have a simple zero there:
$$h(a_1,a_2)=a_1u(a_2). \eqn\gurmo$$
The Hamiltonian vector field generated by this $h$ is
$$V=u(a_2){\partial\over\partial a_2}-{\partial u\over\partial a_2}
a_1{\partial\over\partial a_1}.  \eqn\rudillo$$
Notice that the vector fields of this form generate
reparametrizations of the $a_2$ line and thus obey (part of) a Virasoro
algebra. This precisely corresponds to the Virasoro algebra that we have
been looking at earlier in the conformal field theory.

To determine how such symmetries act on the tachyon field, we compute
$$V(a_1-f(a_2))=-{\partial u\over \partial a_2}a_1-u{\partial f\over
\partial a_2}.\eqn\omo$$
Hence acting by, say, $1+\epsilon V$ transforms
the equation $a_1-f(a_2)=0$ into the equation
$$ a_1-\left(f+\epsilon u{\partial f\over \partial a_2}+\epsilon
{\partial u\over \partial a_2}f\right) = 0  \eqn\opo$$
(up to order $\epsilon^2$).  The transformation law of $f$ is thus
$$\delta f=\epsilon\left(u{\partial f\over\partial a_2}+{\partial u\over
\partial a_2}f\right).\eqn\kopo$$
This is the transformation law of a field of spin 1, in agreement with
what we found in the conformal field theory.

This conclusion could have been obtained without any computation, as follows.
The fact that we are only considering area preserving diffeomorphisms
means that it is natural to think of the $a_1-a_2$ plane as the cotangent
bundle of the $a_2$ line.  The equation $a_1-f(a_2)=0$ thus defines
a section of the cotangent bundle of the $a_2$ line or in other
words a differential form on the $a_2$ line.  Such a differential form
transforms, of course, as a field of spin one.

Had we considered instead Hamiltonians
with a higher order dependence on $a_1$,
say $\widetilde
h=a_1^2v(a_2)$, the same calculation would show that the symmetry generated
by $\widetilde h$ acts nonlinearly on the tachyon field $f$.
In the quantum theory, this means that the symmetries change the number
of particles; for instance, a symmetry transformation might map a
two particle state to a one particle state or vice-versa.

This is a very unusual situation.  In fact, kinematically it is impossible
for massive particles or for massless particles in $D>2$.  The
dispersive propagation of waves would make it impossible for a symmetry to
map an $n$ particle state to an $m$ particle state with $m\not= n$
except for massless particles in $D=2$.

We are, happily, dealing with just that case, so the behavior
just indicated is not impossible, but it is nonetheless very strange from
the point of view of conformal field theory.
Conventionally, in conformal field theory, a symmetry comes from
a world-sheet conserved current $J$.
In quantizing on a cylinder, the conserved charge is defined as
$$A=\oint J,  \eqn\oddo$$
with the integral taken over any contour running once around the cylinder.
Current conservation means that $A$ does not depend on the contour.
$A$ is then a well-defined operator in the one string Hilbert space.
It maps one string states to one string states.  It appears that there
is no room for $A$ to map a one string state to a two string state.
In \S6, we  will  discuss how this can come about.

\section{Discrete Charges on Discrete States}

The purpose of this subsection is to show that the second
term in the conserved charges $\A_{s,n}$, which did not
give a contribution when acting on the tachyon of the
uncompactified theory, is essential to obtain sensible
results for the action on discrete states.
At the same time, we will show explicitly that the new moduli that we have
introduced (corresponding to operators that are not annihilated separately
by $b_0$ and $\overline b_0$) are necessary by showing that they
are created when symmetries act on the ``old'' moduli.

We will consider the simplest example, involving
the discrete charge
$$\A_{{3\over 2},{1\over 2},{1\over 2}} = \CINT {dz \over 2 \pi i}
\, W^+_{{3\over 2},{1\over 2}} \bO_{{1\over 2},{1\over 2}}
- \CINT {d\bz \over 2 \pi i} \, c W^+_{{3\over 2},{1\over 2}}
\bX_{{1\over 2},{1\over 2}},\eqn\discrcharge$$
where $\bX_{{1\over 2},{1\over 2}}= \bar{b} \exp (iX^+)$.

Let us act
%first
on the discrete tachyon:
$$ D_1(w,\bar{w} ) = c(w) W^+_{{1\over 2}, -{1\over 2}}(w)\,
\bar{c}(\bar{w}) \overline{W}^+_{{1\over 2}, -{1\over 2}}(\bar{w}) .
\eqn\disctachi$$
There are no complications in calculating the commutator;
after some work one finds
$$
\{ \A_{{3\over 2},{1\over 2},{1\over 2}} , D_1(w,\bar{w}) \}
= i\sqrt{2} \, c\partial X \, \bar{c} \bar{\partial} X
\,+\, 2c\partial X \,(\bar{\partial} \bar{c} +
{1\over \sqrt{2}} \bar{c} \bar{\partial} \phi )
\, + 2 c \partial c \, \partial X .\eqn\dichondita$$

Let us analyze the result.   The first term on the right
hand side was expected; it corresponds to a closed string
cohomology class at ghost number (1,1), with the corresponding
state annihilated both by $b_0$ and $\bar{b}_0$. This is therefore
a conventional modulus.  The second term on the right hand side
also has ghost number (1,1), but while the holomorphic part
corresponds to a conventional modulus, the antiholomorphic part
corresponds to the operator
$\bar{a}$ discussed in \S2.
This term is BRST invariant, but not annihilated by $b_0-\overline b_0$.
What saves the day is the last term on the right hand side of \dichondita.
This term comes from the second term in the charge, showing that the second
term is necessary.
This last term in \dichondita\ is of ghost number (2,0), and
can be recognized as the BRST invariant operator
$(a \cdot c\partial X)$.
The last two terms in \dichondita\ add up to
$$ 2c\partial X \,(\bar{\partial} \bar{c} +
{1\over \sqrt{2}} \bar{c} \bar{\partial} \phi )
+ 2 c \partial c \, \partial X = (a+ \bar{a} ) \,
(c\partial X \cdot 1 ),\eqn\lasttwo$$
which corresponds to the first ``new'' modulus.

%Our second example consists on the action of the discrete
%charge on the discrete state
%$$D_2(w,\bar{w}) =c\partial X \, \bar{c} \bar{\partial} X
%= Y^+_{1,0} \overline{Y}^+_{1,0}  .
%\eqn\ddii$$

\chapter{Non-Linear Symmetries In Conformal Field Theory}

This final section will be denoted to answering the question
raised at the end of \S5.2 -- how can symmetries that
act nonlinearly on the states
arise in conformal field theory?

To state the answer in the simplest possible form, this can occur
when one has a current $B$ that is not strictly conserved, but conserved
only up to a BRST commutator.  If we think of $B$ as an operator
valued one-form, then
$$ dB^{(1)}=\{Q,B^{(2)}\}, \eqn\unco$$
with $B^{(2)}$ a two-form.

\unco\ leads to a possibility of non-linear symmetry action, as follows.
Recall the conventional derivation of linear Ward identities.
We start with the insertion of $0=dB^{(1)}(z)$ in a correlation
function:
$$0=\int_{z}\langle dB^{(1)}(z)\prod_{i=1}^nT_i(w_i)\rangle.\eqn\hoxo$$
Here $T_i,\,\,\,i=1\dots n$
are some additional $Q$ invariant operators, inserted at points $w_i$.
The symbol $\int_z$ means to integrate in $z$, without integrating over
the $w_i$ (or other moduli of the surface $\Sigma$ on which all this is
happening).  Instead of using the fact that $dB^{(1)}=0$, we
think of $dB^{(1)}$
as an exact differential, and by picking up in the standard way
singularities in the operator products $B^{(1)}(z)T_i(w_i)$, we get
$$0=\sum_{i=1}^n\langle [A,T_i(w_i)]\cdot \prod_{j\not= i}T_j(w_j)\rangle
\eqn\coxo$$
where $A$ is the conserved charge $A=\oint B^{(1)}$.
This is a conventional linear Ward identity.

In the derivation of \coxo, $z$ was the only integration variable,
so the only singularities encountered were singularities in the operator
product of $B$ with a single $T_i$.  This is the reason that the Ward
identity is linear; that is, it involves only terms coming from
$B\cdot T_i\sim T_i'$ with one initial and one final $T$.

\FIG\surfdeg{Linear contributions to Ward identities come from
degenerations of Riemann surfaces of the type indicated in (i),
where a genus zero branch containing a current and precisely one additional
field splits off.  Nonlinear terms come from more slightly more general
degenerations (ii) with a current and more than one additional field
splitting off.  The example given here involves a coupling of a current
to four tachyons.}
We are now interested in a situation in which, instead of
$dB^{(1)}=0$, we have \unco, with some $B^{(2)}$.
The derivation of the Ward identity then begins with
$$\int_{{\cal M}_{g,n+1}}
\langle dB^{(1)}(z)\prod_{i=1}^nT_i(w_i)\rangle
=\int_{{\cal M}_{g,n+1}}
\langle \{Q,B^{(2)}\}
\prod_{i=1}^nT_i(w_i)\rangle. \eqn\mimmo$$
Here instead of just integrating over $z$, we are integrating over
${\cal M}_{g,n+1}$, the moduli space of Riemann surfaces of
genus $g$ with $n+1$ punctures.
Equation
\mimmo\ is an equality between two total derivatives.  The left hand
side has been written in \mimmo\ explicitly as a total derivative
$dB$; the right hand side is implicitly a total derivative since
in string theory an insertion of $\{Q, B^{(2)}\}$
gives a total derivative on moduli space.  (More exactly, it gives a form
that becomes a total derivative after integrating over the position of
insertion of $B^{(2)}$.)

The Ward identity -- the generalization of \coxo\ -- arises now as follows.
Integrating by parts in \mimmo, one picks up surface terms at the various
components of infinity.   The surface terms from
the left hand side of
\mimmo\ are the ones that we have already encountered in the ordinary,
linear Ward identity \coxo.  However, the right hand side of \mimmo\
may contribute additional surface terms.
\foot{These will not involve a coincidence of $B$ with just one other operator,
since the position of insertion of $B^{(2)}$ has already been integrated
to reduce the right hand side of \mimmo\ to a total derivative.}
For instance, as in figure
(\surfdeg(ii)), one component of infinity in ${\cal M}_{g,n+1}$ parametrizes
surfaces which have a genus zero component containing $B$ and two of
the $T$'s.  In the Ward identity, this would be interpreted as a matrix
element $\langle T''|A|T,T'\rangle$ or $\langle T,T'|A|T''\rangle$
depending on whether the $T$'s are positive or negative energy
(incoming or outgoing) states.  ``Non-linear'' terms in the Ward identity
can thus arise from the right hand side of \mimmo.

\section{Invariant Formulation}

We will now restate the forgoing in a more invariant language.
To begin with, obviously a key role was played by the equation \unco.
This equation was, in fact, part of the ``descent'' equations
$$\eqalign{   0 & =\{Q,{\cal O}^{(0)}\} \cr
d{\cal O}^{(0)} & =\{Q,{\cal O}^{(1)}\} \cr
            d{\cal O}^{(1)} & =\{Q,{\cal O}^{(2)}\}.\cr} \eqn\ribbox$$

We can thus give a succinct explanation of where currents come from
and when they act non-linearly.  Currents arise as one-form components
derived from a BRST invariant zero-form observable ${\cal O}^{(0)}$ (in the
semirelative
cohomology);
they act non-linearly when the associated
two-form component is non-vanishing.
For in that case, the right hand side of \mimmo\ must be included.

Now, the discussion of \mimmo\ was rather asymmetrical in several
respects.  We singled out one operator, the ``current,'' for separate
treatment from the other operators $T_i$. This is unnatural since,
just like the $T_i$, the current is derived from a basic zero-form via
the descent equations.  Also, in our discussion of
\mimmo, the exact differentials on the left and right hand sides
appeared to have quite different origins.

We will now give a more symmetric account.  We start with
some BRST invariant operator-valued zero-forms ${\cal O}_a$, $a=1\dots s$,
of ghost number $w_a$ and annihilated by $b_0-\overline b_0$.
It may be that one of the ${\cal O}$'s has ghost number one,
and if so the associated
one-form might be called a
ghost number zero ``current,'' but whether this is so
is immaterial.

Naively speaking, we wish to consider the
``correlation function'' $\Theta=\langle\prod_{a=1}^s {\cal O}_a\rangle$
on a fixed Riemann surface $\Sigma$ with $s$ marked points.
Because of antighost zero modes, $\Theta$ cannot be interpreted as a function
on the moduli space ${\cal M}_{g,s}$; rather it is a differential form.
In fact, the operator formalism, for instance, constructs $\Theta$ as
a {\it closed} differential form of degree
$${\rm deg}\,\,\Theta =6g-6+\sum_{a=1}^s w_a        .\eqn\hicco$$
(This is the number of antighost insertions needed to get a non-zero
result given that the total ghost number inside the correlator must
be $6-6g$.)

Here is a brief sketch of how one proves that $\Theta$ is closed.  For
$\eta$ a Beltrami differential representing a tangent vector to
${\cal M}_{g,s}$ let $\int_\Sigma
\eta b$ be the corresponding
mode of the ghost field $b$.  If the ghost quantum numbers are such that
$\Theta$ is an $n$-form, consider $n+1$ Beltrami differential $\eta_{(i)},
\,\,\,i=1\dots n+1$ .
The $n$ form $\Theta$ is defined by
$$\Theta(\eta_{(1)},\dots,\eta_{(n)})=\langle \prod_{a=1}^s{\cal O}_a
\prod_{j=1}^n\int_\Sigma\eta_{(j)}b\rangle.\eqn\odo$$
Consider the identity
$$0=\langle\prod_{a=1}^s{\cal O}_a\cdot \{Q,
\prod_{j=1}^{n+1}\int_\Sigma \eta_{(i)}b\}\rangle. \eqn\hoddo$$
By use of $\{Q,b\}=T$ ($T$ being here the stress tensor), this takes
the form
$$ 0 = \langle \prod_{a=1}^s{\cal O}_a\sum_{j=1}^{n+1}(-1)^{j-1}
   \int_\Sigma\eta_{(j)}T\prod_{1\leq k\leq n+1,\,k\not= j}\int_\Sigma
\eta_{(k)}b\rangle. \eqn\yoddo$$
Since the insertion of $T$ is a derivative on ${\cal M}_{g,s}$,
the right hand side of \yoddo\ is the antisymmetrized sum of the first
derivatives of $\Theta$, and thus is
$d\Theta(\eta_{(1)},\dots,\eta_{(n+1)})$.
Hence \yoddo\ is the desired result $d\Theta = 0 $.

Now, the case that is most often considered is the case in which
$$\sum_{a=1}^s   w_a = 2s.            \eqn\xulicco$$
In this case, $\Theta$ is a form of degree $6g-g+2s$, \ie, a top
form or measure on moduli space.  In this case, one can integrate
$\Theta$ over ${\cal M}_{g,s}$ to get a number, the
string theory correlation function.  (The integral will converge if there
are no infrared divergences.)

The second most important case is
$$\sum_{a=1}^sw_a=2s-1.        \eqn\ulicco$$
In this case, $\Theta$ is a form of codimension 1.
Since $d\Theta=0$, certainly
$$0=\int_{{\cal M}_{g,s}}d\Theta.  \eqn\licco$$
On the other hand, by Stokes's theorem, if $W_\alpha$ are the components
at infinity in ${\cal M}_{g,s}$, then
$$\int_{{\cal M}_{g,s}}d\Theta=\sum_\alpha \int_{W_\alpha}\Theta.\eqn\icco$$
Combining these, we get the Ward identity
$$ 0 = \sum_\alpha\int_{W_\alpha}\Theta. \eqn\cco$$

The most typical way to obey \ulicco\ is to take one of the ${\cal O}$'s,
say ${\cal O}_1$, to have ghost number one, and the others to have ghost
number two.  Then ${\cal O}_1$ corresponds to a ``current,'' and the
others are ``fields.''  But it is not necessary to have this arrangement.
In general, one might be able to obey
\ulicco\ with no operator having ghost number one (or two).
One would then obtain the Ward identity \cco, but one would have
no temptation to single out one of the operators as a ``current''
and the others as ``fields.'' There is complete symmetry, in the basic
formalism, between the currents and the fields.

\FIG\positstates{Here we show the three tachyons $T_i$ with negative
$X$ momentum, the exceptional tachyon $\widehat T$ and the discrete
state $B$.}

\section{An Example}

\REF\newkleb{I. R. Klebanov, ``Ward Identities In Two-Dimensional
String Theory,'' PUPT-1302 (December, 1991).}
We now want to give a simple example of
the extraction of nonlinear contributions in a Ward identity.\foot{After
we had worked out the conceptual framework explained above
for this calculation,
we received a new paper by Klebanov [\newkleb] in which essentially the
same Ward identity is obtained with a slightly different justification.
Klebanov also works out some consequences
of this Ward identity. Some of the considerations of Kutasov, Martinec,
and Seiberg [\kms] are also closely related.}
We will do this for the uncompactified $D=2$ model, in genus $g=0$,
with the cosmological constant $\mu=0$.
We therefore must recall a few facts about that model.

\REF\grossk{D. J. Gross and I. R. Klebanov, Nucl. Phys. {\bf B359}
(1991) 3.}
\REF\dif{P. Di Francesco and D. Kutasov, Phys. Lett. {\bf 261B} (1991) 385.}
The components of the tachyon vertex operator are
$$\eqalign{
G=2:\quad T^{(0)} & = c\overline c e^{ipX}e^{(\sqrt 2-|p|)\phi} \cr
G=1:\quad T^{(1)} & = \left(dz \,\overline c -d\overline z\,\,c\right)
e^{ipX}e^{(\sqrt 2-|p|)\phi }\cr
G=0:\quad T^{(2)} & = dz\wedge d\overline z\,\, e^{ipX}e^{(\sqrt 2-|p|)\phi
}.\cr
}  \eqn\ruffo$$
Here $p$ is the momentum.  Roughly speaking $p>0$ and $p<0$
correspond to incoming or outgoing states.
It is known [\polyakov,\grossk,\dif]   that
for $g=\mu=0$, the amplitudes $\langle T_{p_1}\dots T_{p_n}\rangle$
vanish unless precisely one $p_i$ is positive or precisely one is negative.
Moreover, the one positive (or one negative) momentum must be an ``exceptional
momentum,'' an integral multiple of $1/\sqrt 2$.

To get a simple, non-trivial example, we will take one tachyon
of positive $X$ momentum $p=1/\sqrt 2$,
and three tachyons of negative momenta $p_i=-q_i,
$ with $q_i>0$, $\sum_{i=1}^3 q_i= \sqrt 2$, and the $q_i$ otherwise
generic.  To keep things clear, we will call the vertex operator
of the positive momentum tachyon $\widehat T$, and we will call the
others $T_i,\,\,\,i=1\dots 3$. Then we have
$T^{(0)}_i = c\bar{c} \, e^{-iq_iX} e^{(\sqrt{2} - q_i)\phi}$ and
$\widehat T^{(0)} = c\bar{c}\, e^{iX/\sqrt{2}} \, e^{\phi/\sqrt{2}}$.

We also will include one other operator, which will be one of the
discrete currents of the $D=2$ model.  In fact, we will pick a current
with $(p_X,p_\phi)=(1,i)/\sqrt 2$.  This operator, call it $B$, has
components
$$\eqalign{ {}&G={}1: \quad B^{(0)}  = \O_{1/2,1/2} \bY_{3/2,1/2} \cr
{}&G={}0: \quad B^{(1)}  = dz\, X_{1/2,1/2} \bY_{3/2,1/2}
+ d\bz \,\O_{1/2,1/2} \bW_{3/2,1/2} \cr
{}&G=-1: \quad B^{(2)}  = dz\,  \wedge \,d\overline z
\,\, X_{1/2,1/2} \bW_{3/2,1/2} .\cr} \eqn\comono$$
Introducing the explicit expressions for the operators we find
$$\eqalign{ B^{(0)} & = \left(cb+{i\over \sqrt 2}\partial(X-i\phi)\right)
\cdot \left(\overline c\left(
(\bar\partial X)^2+{i\over \sqrt 2}\bar\partial^2X
\right)\right)\cdot e^{i(X+i\phi)/\sqrt 2} \cr
B^{(1)} & = \left(dz \,\,b \overline c +d\overline z \,
(cb+{i\over \sqrt 2}\partial(X-i\phi))\right)
\cdot \left((\bar\partial X)^2+{i\over \sqrt 2}\bar\partial^2X
\right)\cdot e^{i(X+i\phi)/\sqrt 2} \cr
B^{(2)} & = dz\,\wedge \,d\overline z
\cdot b\cdot
\left((\bar\partial X)^2+{i\over \sqrt 2}\bar\partial^2X
\right)\cdot e^{i(X+i\phi)/\sqrt 2} .\cr} \eqn\comonoi$$
In the notation of [\witten, eqn. (2.32)], $B^{(1)}$ corresponds to
$\overline J_{3/2,1/2,1/2}$. We will sometimes use the notation
$\Omega^{(1)} = \Omega^{(1)}_z + \Omega^{(1)}_\bz$ where we decompose
any one-form into its $dz$ and $d\bz$ components. The sum of the momenta
of the five operators we are considering are $\sum p_X = 0$
and $\sum p_{\phi} = -i2\sqrt{2}$, as they should be.
Figure \positstates\ shows
the positions of the five relevant operators in the usual diagram
for the states of the theory.

We now want to study the amplitude $\Theta = \langle T^{(0)}_1\,\,
T^{(0)}_2\,\,T^{(0)}_3\,\, B^{(0)}\,\,\widehat T^{(0)}\rangle$.
Here all the operator valued zero forms are of ghost number two
except for $B^{(0)}$ which is of ghost number one. Thus the ghost
numbers add up to nine and we need three antighost insertions to
bring down the ghost number to the standard value of six. Therefore,
as expected $\Theta$ must be a three form, and since the moduli space
of the five-punctured sphere is four-dimensional, $\Theta$ is indeed
a form of codimension one. The Ward identity is to be obtained by
constructing this three form and computing its boundary contributions.

We do the calculation on the complex $z$ plane.  We insert
$T_1^{(0)},T_2^{(0)}$, and $T_3^{(0)}$ at $0,1$, and  $\infty$, the
operator $B^{(0)}$ at $x$ and $\widehat T^{(0)}$ at $y$. The form
$\Theta$ will include three antighost insertions. These can be
represented as $b(v) = \oint b(z) v(z) dz $ where $v(z)$ is the vector
field on the surface that generates, via the Schiffer variation, the
desired change in modulus of the surface (see [\operatorformalism] ).
For example, a change of position of a puncture will be given by a
vector field $v(z) = \epsilon$ (a constant), for which
$b(v) = \epsilon b_{-1}$. But $b_{-1}$ and $\bar{b}_{-1}$ are precisely
the operators that acting on zero form states give the one form
states, and acting on the one form states gives us the two form states
of the descent equations (eq. \statessol ). For our present case, since
all the moduli of the surface can be associated to motions of the
punctures, the antighost insertions sinply turn the zero forms
appearing in $\Theta$ into their descendents. For the fixed tachyons
we use their zero forms $T_i^{(0)}$. The $B$ and $\widehat T$ operators
can appear as the three forms $B^{(1)}(x)\widehat T^{(2)}(y)$
or $B^{(2)}(x)\widehat T^{(1)}(y)$.

The term $B^{(1)}(x)\widehat T^{(2)}(y)$ would have pieces of the
form $dx \wedge dy \wedge d\bar{y}$ and
$d\bar{x} \wedge dy \wedge d\bar{y}$. The first vanishes because
the correlator is zero due to wrong left and right movers ghost
number. The second piece appears at first sight to be
nonzero, but it turns out to vanish because the various possible
contractions of $\left(cb+{i\over \sqrt 2}\partial(X-i\phi)\right)$
with vertex operators and with ghost zero modes add up to zero.
The term $B^{(2)}(x)\widehat T^{(1)}(y)$ would have pieces
$dx\wedge d\bar{x} \wedge dy$ and $dx \wedge d\bar{x} \wedge d\bar{y}$.
The first one vanishes again due to ghost number and the second one
will give us a nonvanishing result.
It follows, then, that the closed three form we are trying to construct
is of the form
$$\eqalign{
\Theta &= q \,\,\,dx\wedge d\overline x \wedge d\overline y \cr
}\eqn\nuff$$
and can be computed by evaluating
$$\langle T_1^{(0)}(0)T_2^{(0)}(1)T_3^{(0)}(\infty)B^{(2)}
(x)\widehat T^{(1)}_{\bar{y}}(y)\rangle .\eqn\trycalc$$
(Notice that not only is $B^{(2)}\not= 0$, making nonlinear
contributions to the Ward identity possible, but with this way of
doing the calculation, all contributions come from $B^{(2)}$.)
The general equation $d\Theta=0$ reduces to
$${\partial q\over \partial y}=0. \eqn\joombo$$

In calculating $q$ various factors must be evaluated.
The Wick contractions of the exponential factors in the vertex operators
give a factor of
$$ |x|^{2-2q_1\sqrt 2}|x-1|^{2-2q_2\sqrt 2}\cdot
{|x-y|^2\over |y|^2|y-1|^2}. \eqn\oombo$$
This certainly does not obey \joombo.  An additional important
factor comes from evaluating the ghost matrix elements.
The left moving ghost fields are $c(0)c(1) c(\infty) b(x)c(y)$, and the
ghost zero mode wave functions are $1,z,z^2$.  Using $c(\infty)$
to absorb the $z^2$ zero mode, and taking all contractions of the
other fields $c(0)c(1)c(y)b(x)$ with each other and with the remaining
ghost zero modes, we find that the amplitude of the left moving ghosts
is  precisely
$${y(y-1)\over x(1-x)(x-y)}.\eqn\hipploz$$
Notice that this precisely cancels the unwanted $y$-dependent
factors in \oombo\ that
violate \joombo.
The right-moving ghost amplitude is just 1.
The remaining factor comes from contractions of the right-moving
oscillator factor $(\bar\partial X)^2+{i\over \sqrt 2}\bar\partial^2 X$
in $\widehat T^{(1)}_{\bar{y}}$.
The sum over contractions gives
$$\left({iq_1\over \overline x}+{iq_2\over \overline x -1}-{i\over \sqrt 2}
{1\over \overline x-\overline y}\right)^2+{1\over\sqrt 2}{q_1\over \overline
x^2}+
{1\over\sqrt 2}{q_2\over (\overline x-1)^2}-
{1\over 2(\overline x-\overline y)^2}.      \eqn\rumpoo$$
Multiplying these factors, the final result is then
$$\eqalign{
\Theta =& dx\wedge d\overline x\wedge d\overline y\cdot
{|x|^{2-2q_1\sqrt 2} \over x} {|x-1|^{2-2q_2\sqrt 2}\over (1-x)}
\cdot {\overline x-\overline y\over \overline y
\cdot (\overline y-1)} \cr &
\cdot \left(
\left({iq_1\over \overline x}+{iq_2\over \overline x -1}-
{i\over \sqrt 2}
{1\over \overline x-\overline y}\right)^2+
{1\over\sqrt 2}{q_1\over \overline x^2}+
{1\over\sqrt 2}{q_2\over (\overline x-1)^2}-
{1\over 2(\overline x-\overline y)^2}\right).  \cr}\eqn\tumpoo$$

Now, we have to extract the boundary terms from the various three
dimensional submanifolds corresponding to degenerations of the moduli
space of the five-punctured sphere .
If ${\cal C}_\epsilon$ is the ball $|x|=\epsilon$ (or $|x-1|=\epsilon$,
or $|1/x|=\epsilon$) with $y$ unconstrained, then
$$\int_{{\cal C}_\epsilon}\Theta = 0, \eqn\cumpoo$$
just because when restricted to ${\cal C}_\epsilon$, $dx$ and
$d\overline x$ are proportional (and $dy$ is missing).
This is not surprising
since such terms would correspond to linear terms in the Ward
identity where $B$ meets one of the
tachyons $T_i$ of negative $X$ momentum. Such terms should vanish
since there is no cohomology at the corresponding total momentum.
Nonvanishing terms arise for $y$ near $a=0,1,\infty$, or $x$.
Notice that in each case, for $y$ near $a$,
$\Theta$ is proportional to $d\overline y/
 (\overline y -\overline a)$ (or $d\overline y/\overline y$ for $a=\infty$).
The surface term near $y =a$ is just
the residue of $\Theta$, that is the coefficient of $d\overline y/
(\overline y-\overline a)$.
The Ward identity is
$$ 0=\int_x\sum_{a=0,1,\infty,x} {\rm Res}_{y=a}\Theta. \eqn\forfo$$
The symbol $\int_x$ refers to an integration over the complex $x$ plane;
the residue of $\Theta$ is a two
form proportional to $dx\wedge d\overline x$ which can be so integrated.

The term in \forfo\ with $a=x$ gives an ordinary linear term in the
Ward identity, of the form $A|\widehat T\rangle\sim|\widehat T'\rangle$,
where $A=\oint B^{(1)}$ is the conserved charge and $\widehat T'$ is
a tachyon state of exceptional $X$ momentum $+\sqrt 2$.

More interesting for our present purposes are the terms with $a=0,1,\infty$.
These arise in the above computation from short distance singularities
of $\widehat T$ with one of the $T_i$.
Such singularities arise in the region of moduli space near
a degeneration of the type in figure (\surfdeg(ii)) in
which $\widehat T$ and one of the $T_i$ are in one branch,
say $\Sigma_1$, while
$B$ and the other two $T_i$ are contained in the other branch $\Sigma_2$.
In the spirit of the conventional linear Ward identity \coxo, the
effect of the branch $\Sigma_2$ as seen by an observer on $\Sigma_1$
is
to insert on $\Sigma_1$ a new operator at the point $P$.
In the problem at hand, this operator turns out to be a tachyon operator,
so the two tachyon operators on $\Sigma_2$ are effectively replaced by one.
The interpretation of this in terms of matrix elements of the conserved
charge must therefore involve terms in which the number of
tachyons is not conserved. For instance, the term with $a=\infty$
corresponds to a process
$$A\,|\,T^{(0)}_1(p_1)\, T^{(0)}_2(p_2)\,\rangle
=\lambda \,\,|\, T^{(0)}(p_1+p_2+1/\sqrt 2)\rangle ,\eqn\mccv$$
in which $A$ maps two tachyons of $X$ momentum $p_1$ and $p_2$ to
a tachyon of momentum $p_1+p_2+1/\sqrt 2$ with an amplitude $\lambda$.
(The Liouville momentum automatically works out correctly in this process.)
Thus, this  operator $A$ has matrix elements
that reduce by one the number of tachyons of negative $p_X$.
Indeed, by considering $p_X$ and $p_\phi$, one can easily
show that the only non-zero matrix elements of
$A$, when acting on tachyons of negative $p_X$,
are those that reduce the number of tachyons by one.
This is the justification for writing \mccv\ as we have,
and not just as a much weaker
statement $\langle T(p_1+p_2+1/\sqrt 2)|A|T(p_1)T(p_2)\rangle=\lambda$.

%As for the value of the matrix element $\lambda$, it is simply the integral
%of the residue
%of $\Theta$  at $y=\infty$, or
%$$\eqalign{\lambda =&
%\int dx\wedge d\overline x \,\,\,{|x|^{2-2q_1\sqrt 2}\over x}
%{|1-x|^{2-2q_2\sqrt 2}\over (1-x)} \cr &\cdot\left(
%\left({iq_1\over \overline x}+{iq_2\over \overline x -1 }\right)^2
%+{1\over \sqrt 2}{q_1\over \overline x^2}+{1\over \sqrt 2}{q_2\over
%(\overline x-1)^2}\right).   \cr}  \eqn\rdox$$
%This integral is readily verified to be the correlator
%$$\langle\!\langle (\partial c+\overline\partial\overline c)T^{(0)}(p_3)
%\,T^{(0)}_1 \,T^{(0)}_2 \, B^{(2)}
%\rangle\!\rangle,\eqn\cuxu$$
%where $p_3 = -p_1 -p_2 -1/\sqrt{2}$.

We will now give a
theoretical explanation for how this arises and how to determine
the value of $\lambda$.

Let $\epsilon$ be a measure of the distance from the degeneration
at $a=\infty$.  At $\epsilon=0$, the Riemann sphere breaks up into
the two components $\Sigma_1$ and $\Sigma_2$, joined at a double
point $P$.  In addition to $P$, $\Sigma_1$ contains
$\widehat T^{(0)}$ and
$T^{(0)}_3$, and $\Sigma_2$ contains the other three operators.
$\Sigma_1$ has the moduli space ${\cal M}_{0,3}$ consisting of one point,
and $\Sigma_2$ has a two dimensional moduli
space ${\cal M}_{0,4}$.
Superficially, factorization of physical states at $P$ appears to involve
the sum over states $|{\cal O}^a\rangle \langle {\cal O}_a |$,
where ${\cal O}^a$ is the dual state to ${\cal O}_a$ (so on the sphere
$\langle {\cal O}^a{\cal O}_b\rangle =\delta^a_b$).
However, the computation determining $\lambda$ involved integrating
$\Theta$ not over ${\cal M}_{0,3}\times
{\cal M}_{0,4}$ but over the cycle ${\cal C}_\epsilon$
in ${\cal M}_{0,5}$.
The third dimension in ${\cal C}_\epsilon$ (relative to ${\cal M}_{0,3}
\times {\cal M}_{0,4}$ over which it is fibered) is the twist angle
joining $\Sigma_1$ and $\Sigma_2$.  In the above computation, extraction
of the residue at $y=\infty$ was a way of integrating over the twist
angle and reducing the integral over ${\cal C}_\epsilon$ to an integral
over ${\cal M}_{0,3}\times {\cal M}_{0,4}$.
As for the three Beltrami differentials on ${\cal C}_\epsilon$, two of
them are the ones we want for eventually integrating on ${\cal M}_{0,3}
\times {\cal M}_{0,4}$.  The third one, associated with the twist angle,
is an insertion of $b_0-\overline b_0$ on the long neck which near
$\epsilon=0$ joins $\Sigma_1$ and $\Sigma_2$.  Because of this insertion,
the sum over physical states arising at the double point $P$ is not
the naive $\sum_a|{\cal O}^a\rangle\langle{\cal O}_a|$
but
$\sum_a|\widetilde{\cal O}^a\rangle\langle{\cal O}_a|  $
where $\widetilde |{\cal O}^a\rangle
=(b_0-\overline b_0)|{\cal O}^a\rangle$.
In later work, we will restrict the symbol ${\cal O}_a$ to run
over plus states only (and not all states as above), and then the sum
over states at a degeneration is
$$ \sum_a|{\cal O}_a\rangle\langle\widetilde{\cal O}^a|
+\sum_a|\widetilde{\cal O}^a\rangle\langle{\cal O}_a|. \eqn\opxx$$

We therefore have that
$$ \int_x {\rm Res}_{y=\infty}\Theta =
\sum_a \, \langle
\!\langle \,\widehat T^{(0)} \, T^{(0)}_3\, \widetilde \O^a
\,\rangle \!\rangle\,\,
\langle\!\langle \,\O_a \,\,T^{(0)}_1 \,T^{(0)}_2 \, B^{(2)}
\rangle\!\rangle .\eqn\factward$$
where the symbol $\langle\!\langle \cdots \rangle\!\rangle$ indicates
that the relevant integral over moduli space should be done
(the object inside such correlator must have the right ghost
number to be a top form).
The states appearing at $P$ are BRST invariant, since they appear
in the operator product of the BRST invariant operators on $\Sigma_1
$ or $\Sigma_2$, and can be considered to represent BRST cohomology
classes, since BRST trivial states would decouple.
(If we were considering loops, BRST non-invariant states would circulate
around the loops and cancel in pairs.)
It is natural to define the nonlinear action of the charge $A=\oint B$ by
saying that for $T^{(0)}_i(p_i)$ states on the incoming branch
$$A\,|\,T^{(0)}_1(p_1)\, T^{(0)}_2(p_2)\,\rangle
=\sum_a \lambda_a \,\,
|\widetilde \O^a\rangle , \eqn\mccvv$$
where $\lambda_a \equiv \langle\!\langle \O_a \,T^{(0)}_1 \,T^{(0)}_2 \,
B^{(2)}
\rangle\!\rangle$.  A similar formula, with kets replaced by bras,
holds for $T^{(0)}_i(p_i)$
on the outgoing branch.  (It is no accident
that the nonlinear term arises only when $p_1$ and $p_2$ are both on
the same branch.)

For our present case the only contribution will come from
the (zero form) tachyon state $\widetilde \O^a = T^{(0)}(-p_3)$,
where $p_3 = -p_1 -p_2 -1/\sqrt{2}$.
Then $\O_a = (\partial c+\overline\partial\overline c)T^{(0)}(p_3)$,
and therefore equation \mccvv\ just reduces to \mccv. The value of
$\lambda$ is therefore defined as
$$\lambda = \langle\!\langle \,
(\partial c+\overline\partial\overline c)T^{(0)}(p_3)
\,T^{(0)}_1 \,T^{(0)}_2 \, B^{(2)}
\rangle\!\rangle.\eqn\valuelambda$$
One can use directly equation \factward\ to get
$$\eqalign{
\lambda \cdot \langle
\!\langle \,\widehat T^{(0)} \, T^{(0)}_3\, T^{(0)}(-p_3)
\,\rangle \!\rangle &= \int_x {\rm Res}_{y=\infty}\Theta \cr
{}&= \int dx\wedge d\overline x \,\,\,{|x|^{2-2q_1\sqrt 2}\over x}
{|1-x|^{2-2q_2\sqrt 2}\over (1-x)} \cr
{}&\cdot\left(
\left({iq_1\over \overline x}+{iq_2\over \overline x -1 }\right)^2
+{1\over \sqrt 2}{q_1\over \overline x^2}+{1\over \sqrt 2}{q_2\over
(\overline x-1)^2}\right).\cr}\eqn\bottomline$$
This concludes our explanation on the origin of the nonlinear
action of the symmetry charges.

\section{Homotopy Lie Algebras}

\REF\stasheffi{J. D. Stasheff, ``{Towards a closed string field
theory: topology and convolution algebra},'' preprint UNC-MATH-90/1.}
One may ask what happens to the Jacobi identity when there are nonlinear
terms in the Ward identities of currents.
Recently,
Kontsevich [\kontsevich]
has used in Chern-Simons gauge theory the notion of homotopy Lie algebras,
whose origins are in old work of Stasheff [\stasheff].
\foot{Kontsevich  showed that it is
possible to do Chern-Simons gauge theory using homotopy Lie algebras instead
of ordinary ones.  In that application he considers only algebras preserving
a metric, which we will not insist on.
Homotopy Lie algebras are closely related to
homotopy associative algebras, which have been suggested
by some authors [\stasheffi ] to be relevant to string field
theory and have been used by Kontsevich
in roughly that connection.}
We will now argue that the nonlinear terms in the Ward
identities generate a homotopy Lie algebra.
\foot{While we were checking the axioms of such an algebra,
we learned of parallel ideas by E. Verlinde that in some respects go farther.}

Given a graded Lie algebra with generators $T_a$ (which may be even or odd),
introduce objects $\eta^a$ with dual quantum numbers and opposite statistics.
(In our application, the reversal of statistics comes about
because the symmetry generators
have the statistics of the currents or one form components of operators,
while the two form components have opposite statistics.)
Introduce a vector field
$$ V=f_{a_1a_2}^b\eta^{a_1}\eta^{a_2}{\partial\over\partial \eta^b}
 +f_{a_1a_2a_3}^b\eta^{a_1}\eta^{a_2}\eta^{a_3}{\partial\over\partial \eta^b}
 +f_{a_1a_2a_3a_4}^b\eta^{a_1}\eta^{a_2}\eta^{a_3}\eta^{a_4}{\partial
\over\partial \eta^b}+\dots . \eqn\gimo$$
The $f$'s are $c$-number ``structure constants.''
$V$ is required to be odd, so if the $T_a$ are all bosonic,
and the $\eta^a$ hence all fermionic,
then $f_{a_1a_2\dots a_n}^b$ vanishes for odd $n$.
On $V$ impose the equation
$$ 0 = \{V,V\}.      \eqn\imo$$
The term in \imo\ cubic in $\eta$ is the conventional Jacobi identity
for the ordinary structure constant
$f_{a_1a_2}^b$.  (This would no longer be so if one permits $V$ to contain
terms of less than quadratic order in $\eta$.)
\imo\ defines a ``homotopy Lie algebra.''
We want to show how to formally extract such a homotopy Lie
algebra from $D=2$ string theory.  In doing so, we will consider as
generators just the plus
states, though the minus states could be included.

Using the notation introduced in the previous subsection, for every
BRST cohomology class ${\cal O}_a$ of plus states, there is a dual
minus state ${\cal O}^a$. If ${\cal O}_a$ is annihilated by
$b_0-\overline b_0$, then ${\cal O}^a$ is not. Again, $\widetilde{\cal O}^a
=(b_0-\overline b_0){\cal O}^a$.
When one derives Ward identities, one picks up contributions when a surface
$\Sigma$ degenerates.  We want to consider the contributions when several
currents collide; then $\Sigma$ splits (as in figure (\surfdeg)) to two
components $\Sigma_L$ and $\Sigma_R$, sharing a point $P$.  Also, $\Sigma_L$
has genus zero.  The operators appearing at $P$ on the two sides are BRST
invariant (since they are produced by ``integrating out'' the BRST invariant
objects on $\Sigma_L$ or $\Sigma_R$) and annihilated by $b_0-\overline b_0$
(since insertions of other operators do not make sense)
and can be considered as BRST cohomology
classes (since a zero form operator $\{Q,\dots\}$ would decouple).
The sum over cohomology classes appearing at $P$ is
$$\sum_a  |{\cal O}_a\rangle\langle\widetilde{\cal O}^a|
+\sum_a |\widetilde {\cal O}^a\rangle\langle{\cal O}_a|, \eqn\jimo$$
as was explained in \opxx.

Define now
$$f_{a_1a_2\dots a_n}^b=\langle\!\langle
{\cal O}_{a_1}{\cal O}_{a_2}\dots
{\cal O}_{a_n}\widetilde {\cal O}^b \rangle\!\rangle ,\eqn\docc$$
where $n\geq 2$ and all the $\O$'s are BRST classes (zero forms).
Here $\langle\!\langle  \dots \rangle\!\rangle$
is a genus zero correlation function
(integrated over the moduli of the $n+1$ points in the standard way).
We claim that if $V$ is defined with these $f$'s, then
the axiom $\{V,V\}=0$ of a homotopy Lie algebra is obeyed.

To prove this, note that $f_{a_1a_2\dots a_n}^b$ as defined in \docc\
will vanish unless the sum of the ghost numbers of the indicated states
is $2n+2$ ($=\hbox{dim}\,({\cal M}_{0,n+1}) + 6$ ).
To derive a Ward identity, one considers the case in which
the sum of the ghost numbers is $2n+1$, so that the ``correlation function''
$\Theta = \langle {\cal O}_{a_1}{\cal O}_{a_2}\dots
{\cal O}_{a_n}\widetilde {\cal O}^b \rangle$
is a closed differential form of codimension one.
The Ward identity is then as usual
$$\sum_\alpha \int_{{\cal M}_\alpha}\Theta = 0 \eqn\hoxo$$
where $\alpha$ labels the possible ways that the surface $\Sigma$ can
degenerate; these are in one to one correspondence with divisions
of the set $\{a_1,\dots ,a_n\}$ into two disjoint
subsets $\{u_1,\dots, u_{n_\alpha}\}$
and $\{v_1,\dots ,v_{n-n_\alpha}\}$.  The Ward identity is concretely
$$0=\sum_\alpha
\langle\!\langle {\cal O}_{u_1}{\cal O}_{u_2}\dots {\cal O}_{u_{n_\alpha}}
\widetilde{\cal O}^a
\rangle\!\rangle \cdot
\langle\!\langle {\cal O}_a{\cal O}_{v_1}\dots
{\cal O}_{v_{n-n_\alpha}}\widetilde{\cal O}^b
\rangle\!\rangle,  \eqn\hcoc$$
where we have used \jimo\ and the fact that by Liouville momentum counting,
there must be at least one ``minus'' operator on each side to get a nonzero
result.
\foot{Were it not for this fact, we would have had to include all of the
operators as symmetry generators, which might be natural anyway.}
Expressed in terms of the $f$'s, \hcoc\ is the component
of the relation $\{V,V\}=0$ proportional to $\eta^{a_1}\dots\eta^{a_n}$,
and thus we have justified that relation.

\section{Application In $D=4$?}

Finally, one might wonder any of these issues could be relevant in a realistic
string theory with a macroscopic four dimensional target space.

The conventional Poincar\'e symmetries in string theory are
derived from operators such as $c\partial X^\mu$ whose
two-form components vanish.
That is why they act linearly.
However, one can at least imagine that there might exist a four dimensional
Poincar\'e invariant vacuum state, with suitable Poincar\'e invariant
couplings between ``matter fields'' and ``ghosts'' (once these are coupled
the distinction between them is fuzzier) such that the Poincar\'e
currents would come from operators with non-zero two-form components.

Even if this occurs, the Poincar\'e symmetries
will act linearly on the particles of the theory, this being inevitable
kinematically in $D>2$.  However, it is at least conceivable that the
Poincar\'e currents could act nonlinearly on themselves.
If the novel terms are entirely of higher than the usual order,
such a theory would be much like an ordinary Poincar\'e invariant
theory, but with some unusual higher order ``anomalous'' Ward identities
presumably involving the spacetime stress tensor.
One might speculate that such identities could shed light on the vanishing
of the cosmological constant.

\chapter{Some Additional Comments}

\REF\DasJevicki{S. R. Das and A. Jevicki, Mod. Phys. Lett.
{\bf A5} (1990) 1639.}
\REF\Sen{A. Sen, Nucl. Phys. {\bf B345} (1990) 551;
{\bf B347} (1990) 270.}
\REF\zwiebach{B. Zwiebach, ``Recursion Relations in Closed String
Field Theory,'' Proceedings of the ``Strings 90'' Superstring
Workshop. Eds. R. Arnowitt, et.al. (World Scientific, 1991)
pp. 266-275.}
\REF\wiesbrock{H. W. Wiesbrock, ``{The Construction of the
Sh-Lie-Algebra of Closed Bosonic Strings},'' FUB-HEP-90 (1990).}
One of the important questions within two dimensional string
theory is that of formulating a string field theory capable of
describing the various possible backgrounds.
On the one hand we have the collective string field
theory [\DasJevicki ] which affords a very successful description
of the tachyon dynamics, but by not incorporating explicitly the
degrees of freedom corresponding to the discrete states makes
it difficult to study the changes of backgrounds. On the other
extreme, we have the BRST closed string field theory, which,
by construction describes correctly the perturbative dynamics
of the theory to all orders. A background is needed to define it,
but it is at least formally background
independent: classical solutions shift backgrounds [\Sen ].
In this string field theory there is not only a tachyon field
but also a field for every discrete state, and infinitely
more fields one for each Fock space state in the Hilbert space.

For a solvable theory such as the two-dimensional string one
could perhaps expect that a suitable truncation of the field
content of the BRST string field theory may afford a complete
and elegant formulation, which could even be completely
background independent, in the sense that no specific background
would be necessary to define the theory.
It is clear that the discrete states we have considered must
be included along with the tachyon to get a complete description.
However, at discrete radii other than the $SU(2)$ point, other
discrete states would appear.  Perhaps they should be included also;
an overall framework for doing this is not evident.

%What we have found
%in this work is that the minimal content, defined by
%the conventional discrete {\it states} is
%not sufficient. The symmetry structure uncovered in [\witten ]
%requires that we include all states corresponding to the
%semi-relative closed string cohomology, the
%cohomology relevant in BRST closed string field theory. It is
%fairly clear, however, that we have not yet obtained a
%background independent formulation. Possibly, we still
%have to add more degrees of freedom, or alternatively,
%change somewhat the rules of the game. One obvious candidate
%for extra degrees of freedom consists of introducing a
%(two-dimensional) field for each discrete state. This field
%content is still exponentially smaller that that of the
%BRST string field theory. There is, of course, no guarantee that
%these additional degrees of freedom will suffice for a satisfactory
%formulation.

Another important issue is that of the spacetime interpretation
of the model. It was proposed in [\witten ] that the three-
dimensional cone $Q$ plays the role of spacetime in the
compactified model.
We have indeed succeeded in describing many phenomena in terms of the
differential geometry of $Q$.
Hopefully, some generalization of this differential geometric setup
will survive when the appropriate additional degrees of freedom
are incorporated.

%We have found here many indications supporting this
%statement. While this cone arises as the manifold where the BRST
%classes at ghost number zero play the role of (polynomial)
%functions, we have found that the semirelative
%cohomology classes having to do with states and symmetries
%fall nicely into differential forms in this cone. Moreover,
%the crucial $b_0 -\bar{b}_0$ operator, which is not a local
%operator on the world sheet, becomes simply the exterior
%derivative, a local operator on the cone. One is able now
%to write the three-point couplings of the theory as an
%abelian gauge theory on a four-dimensional space closely
%related to the cone. In some sense, however,
%the cone seems to play the role of an internal space.
%As we mentioned above, presumably one must still enlarge the
%space of fields, and in that case letting the states become
%fields corresponds to adding the two extra dimensions corresponding
%to the conventional space and time of the model. This
%suggests that the proper description could be six-dimensional.

Throughout this paper we have had to deal with subtleties
associated with the fact that many relations hold only up
to BRST trivial terms. A very similar phenomenon is familiar
in BRST closed string field theory [\zwiebach ]. The nonpolynomial
structure of the theory arises because the failure of the
string products to associate (more precisely, to
give Jacobi identities) is repaired by the BRST operator
acting on the higher order string products. This type of
structure is called a homotopy Lie algebra [\wiesbrock ,\stasheff ],
the higher string
products defining the homotopies. The nonlinear symmetries,
studied here seem to define a similar structure. It is of
great interest to understand the relations clearly, as
this may give us the tools to investigate concretely
the enormous symmetry structure of critical
closed string theory.
\bigskip

\ack
We are indebted to E. Verlinde for stimulating discussions.

\APPENDIX{A}{Schur Polynomials and Explicit Representatives}

\medskip
\noindent
$\underline{\hbox{ Schur Polynomials}}$.
The elementary Schur polynomials $S_k$ with $k=0,1,2,\cdots $, are
defined via the following generating function:
$$\sum_{k=0}^\infty S_k(x_j) z^k = \exp \left(
\sum_{k=1}^\infty x_k z^k \right),\eqn\defschur$$
where the $x_k$'s , with $k=1,2,\cdots$, are the arguments of the polynomials.
The first few examples are:
$$S_0 = 1,\quad S_1 = x_1,\quad S_2 = x_2 + {1\over 2} x_1^2,\quad
S_3 = x_3 + x_2x_1 + {1\over 3!} x_1^3.\eqn\exchur$$
The Schur polynomials can be reduced to factorials under some
circumstances. If
$$x_k = {(-1)^k \over k} a \equiv c_k a ,\eqn\speccase$$
use of the generating function above gives
$$\sum_{k=1}^\infty S_k(\vec{c} a) z^k
= \exp \left( -a \sum_{k=1}^\infty
{(-1)^{k+1} \over k} z^k \right)
= \exp \left( -a \ln (1+z) \right) = (1+z)^{-a},\eqn\simplschur$$
where from we deduce
$$S_k(\vec{c} a) = {1\over k!} \left( {d\over dz}\right)^k
(1+z)^{-a} \vert_{z=0} = {(-1)^k \over k!} \,
{\Gamma (a+k) \over \Gamma (a)}.\eqn\fsimpsch$$
Another simple consequence of the generating function is the
identity
$$S_k (x+y) = \sum_{j=0}^k S_j(x) S_{k-j} (y).\eqn\addschur$$
Using this and \fsimpsch\ one finds that
$$S_k[\vec{c}\, (\sqrt{2}-1)] = S_k (\vec{c}\,\sqrt{2}p)
+ S_{k-1} (\vec{c}\sqrt{2}p),\eqn\propschi$$
and this implies that
$$\sum_{q=0}^{n} (-)^q S_{n-q} (\vec{c}\,[\sqrt{2}p -1])
= S_{n}(\vec{c}\sqrt{2}p).\eqn\propschii$$
\medskip
\noindent
$\underline{\hbox{Ground Ring Representatives}}$.
The state corresponding to the ground ring operator
$\O_{k,k}$ ($k>0$) was given by Bouwknegt et. al. and it reads
\footnote{*}{We have changed the overall sign of the
representative, for convenience.}
$$\ket{\O_{k,k}} = -\sum_{q=1}^{2k+1} S_{2k+1-q}
\left({-\alpha_{-j}^-\over j}\right) \, b_{-q}
\ket{\sqrt{2}k , i\sqrt{2}k ; 0} \eqn\repii$$
where $\alpha_{-j}^- \equiv {1\over \sqrt{2}} (\alpha_{-j}
+ i\phi_{-j})$ and the vacuum $\ket{p_X,p_\phi;0}$ is taken
to be equal to $c_1\ket{p_X,p_\phi ;{\bf 1}}$.
The first two examples are
$$\ket{\O_{0,0}} = b_{-1}\ket{0} = \ket{{\bf 1}}$$
$$\ket{\O_{{1\over 2},{1\over 2}}}
= \left( c_1b_{-2} + {1\over \sqrt{2}}
(\alpha_{-1} -i\phi_{-1}) \right) \ket{ 1/ \sqrt{2},
i/\sqrt{2} ; {\bf 1}}$$
It is clear that the states $\ket{\O_{k,k}}$ satisfy the
conditions
$$\{b_0 , b_1, b_2, \cdots \} \ket{\O_{k,k}} = 0,\eqn\condlemm$$
and therefore the above representatives are automatically
highest weight states (and define the primary operators $\O_{k,k}$).
The above conditions, as shown in the Lemma (sect.4), imply
that
$$\ket{\partial \O_{k,k}} = Q \ket{X_{k,k}},\quad
\hbox{with} \quad \ket{X_{k,k}} = b_{-1} \ket{\O_{k,k}}.\eqn\intrx$$
The states $\ket{X_{k,k}}$ are therefore given by
$$\ket{X_{k,k}} = \sum_{q=1}^{2k} S_{2k-q}
({-\alpha_{-j}^-\over j}) \, b_{-(q+1)}b_{-1}
\ket{ \sqrt{2}k , i\sqrt{2}k ; 0} \eqn\repxii$$
The first two examples for these states are
$$\ket{X_{0,0}}= 0,\quad \ket{X_{{1\over 2},{1\over 2}}} =
b_{-2}b_{-1} \ket{ 1/ \sqrt{2}, i/ \sqrt{2} ; 0}
= b_{-2} \ket{ 1/ \sqrt{2}, i/ \sqrt{2} ; {\bf 1}}.
\eqn\examplx$$

The operators $\O_{k,k}$ and $X_{k,k}$ corresponding to the
states quoted above are simply given by
$$\O_{k,k}= \left( -S_{2k}({-i\over j!} \partial^j
X^-) + \sum_{q=1}^{2k} S_{2k-q}({-i\over j!} \partial^j
X^-)\, c\, { \partial^{q-1}\,b \over (q-1)!}\, \right)
e^{i2k\,\bX^+}\eqn\repcoh$$
$$X_{k,k}= \sum_{q=1}^{2k} S_{2k-q}({-i\over j!} \partial^j X^-)
\, { \partial^{q-1}\, b \over (q-1)!}   \,
e^{i2k\,\bX^+}\eqn\repcurr$$
The first examples for these operators are
$$\eqalign{
&\O_{0,0} = 1 ,\quad \O_{{1\over 2},{1\over 2}} =
(cb + i\partial X^-)\exp(iX^+),\cr
& X_{0,0} = 0,\quad X_{{1\over 2},{1\over 2}} =
b\exp(iX^+).\cr}\eqn\lastexp$$
One final identity concerns the operator product expansion
of an antighost with the $\O$'s:
$$b(z)\, \O_{k,k}(w) = {1\over z-w} X_{k,k}(w) + \, \hbox{regular}.
\eqn\opebx$$

\refout
\figout
\end